\documentclass[letterpaper, 11pt]{article}
\pdfoutput = 1

\usepackage{shortcuts}

\usepackage[margin = 2.5cm]{geometry}
\usepackage{tikz}
\usetikzlibrary{arrows, decorations.pathmorphing, decorations.markings, backgrounds, positioning, fit, petri}
\usepackage[outline]{contour}
\contourlength{1.2pt}

\begin{document}

\thispagestyle{empty}

\begin{flushright}
\texttt{BRX-TH-6657}
\end{flushright}

\begin{center}

\vspace*{5em}

{\LARGE \bf Systematic Constructions of Fracton Theories}

\vspace{1cm}

{\large \DJ or\dj e Radi\v cevi\'c}
\vspace{1em}

{\it Perimeter Institute for Theoretical Physics, Waterloo, Ontario, Canada N2L 2Y5}\\
{\it Martin Fisher School of Physics, Brandeis University, Waltham, MA 02453, USA}\\
\texttt{djordje@brandeis.edu}\\

\vspace{0.08\textheight}
\begin{abstract}
{\normalsize Fracton theories possess exponentially degenerate ground states, excitations with restricted mobility, and nontopological higher-form symmetries. This paper shows that such theories can be defined on arbitrary spatial lattices in three dimensions. The key element of this construction is a generalization of higher-form gauge theories to so-called $\Ff_p$ gauge theories, in which gauge transformations of rank-$k$ fields are specified by rank-$(k - p)$ gauge parameters. The $\Z_2$ rank-two theory of type $\Ff_2$, placed on a cubic lattice and coupled to scalar matter, is shown to have a topological phase exactly dual to the well-known X-cube model. Generalizations of this example yield novel fracton theories. In the continuum, the U(1) rank-two theory of type $\Ff_2$ is shown to have a perturbatively gapless fracton regime that cannot be consistently interpreted as a tensor gauge theory of any kind. The compact scalar fields that naturally couple to this $\Ff_2$ theory also show gapless fracton behavior; on a cubic lattice they have a conserved U(1) charge and dipole moment, but these particular charges are not necessarily conserved on more general lattices. The construction straightforwardly generalizes to $\Ff_2$ theories of nonabelian rank-two gauge fields, giving first examples of pure nonabelian higher-rank theories.}
\end{abstract}
\end{center}

\newpage

\tableofcontents

\newpage

\section{Introduction}

\subsection{Fractons}

Recent years have seen a surge of interest in the physics of fractons.\footnote{See the review \cite{Nandkishore:2018sel} for an exhaustive list of references. Papers particularly relevant to this work are \cite{Haah_2011, Vijay:2016phm, Pretko:2016kxt, Halasz:2017, Shirley:2017suz, Bulmash:2018lid, Ma:2018nhd, Song:2018gbb, Bulmash:2018knk, Williamson:2018ofu, Weinstein:2018xil}.} Given the diversity of examples being discussed, it may be too early to simply say what a fracton \emph{is}, but nevertheless it should be fair to say that a quantum theory of fractons has the following extraordinary features:
\begin{enumerate}
  \item[F1.] The theory, which can be either gapped or gapless, has a ground state degeneracy that scales exponentially with the (linear) size of the system.
  \item[F2.] The particle excitations possess restricted mobility. For example, some particles may only move along certain directions in space.
  \item[F3.] The theory has nontopological higher-form symmetries \cite{Paramekanti:2002, Xu:2007, Nussinov:2006iva, Nussinov:2009zz}. For example, there may exist an independent conserved quantity associated to each $x$-, $y$-, or $z$-plane in a 3d space.
\end{enumerate}

The optimal definitions of these features are still unsettled. For instance, there are many ways to have ``restricted mobility'' in F2 (two wide classes of mobility restrictions were formulated in Refs.~\cite{Haah_2011, Vijay:2016phm}). It is also natural to supplement F1 by demanding that a gapped fracton theory have a topologically protected ground state degeneracy. This requirement will not be imposed in this paper; gapped and gapless theories will be treated on the same footing.

All known examples that satisfy the desiderata F1--3 in any reasonable sense live on spaces that have at least some special structure. In a discrete setup, fracton theories are defined on spatial lattices that are topologically cubic or that can be decomposed into subdimensional layers \cite{Shirley:2017suz, Slagle:2018swq}. There is an analogous limitation in the continuum, where known examples of fractons are certain tensor gauge theories \cite{Pretko:2016kxt} that can be defined on manifolds, i.e.~on locally Euclidean (possibly curved) spaces, but not on more general topological spaces.

The goal of this paper is to show that fracton theories may in fact be defined on arbitrary spatial lattices. This will be done constructively, by taking the simplest known fracton theory on a cubic lattice --- the X-cube model \cite{Vijay:2016phm} --- and by explicitly defining its natural generalization that can be placed on \emph{any} cellulation of a three-dimensional space. This generalization crucially involves a novel higher-rank gauge theory of a quite unusual yet surprisingly natural sort. One of the intriguing features of this gauge theory is that it can be defined for \emph{any} gauge group --- even a nonabelian one. (The most natural higher-rank lattice gauge theories, described in the continuum by higher-form fields, do not have nonabelian generalizations, unless they are coupled to additional fields that equip them with higher group symmetries \cite{Baez:2002jn}.) For gauge group $\Z_2$, if this new gauge theory is placed on a cubic lattice, coupled to matter, and taken to its topological/weak coupling limit, the resulting theory becomes \emph{exactly} dual to the X-cube model. The fracton features F1--3, however, can be found even without matter, on arbitrary lattices, and with arbitrary gauge groups (at least at energies above any confinement scale). The paper will work in 3d, but the analysis will generalize to other dimensions.

The main result of this paper is certainly the very definition of this new class of gauge theories. They will be called $\Ff_p$ theories, and a telegraphic overview of their definition will be given in the next Subsection. To whet the reader's appetite, here are three important lessons that can be learned about fractons by moving away from situations with cubic symmetry:
\begin{enumerate}
  \item Not all properties of hitherto studied fracton theories are needed to satisfy requirements F1--3; some are merely accidents due to the cubic symmetry. A salient example is the conservation of dipole moment, whose existence and gauging are often associated to fractons in the literature \cite{Pretko:2016kxt}. As a counterexample, this paper will explicitly show that fracton matter on a body-centered-cubic lattice need not have either conserved charge or conserved dipole moment, while exhibiting other kinds of subdimensional symmetries, exponential topological degeneracy, and excitations with restricted mobility.
  \item The natural continuum examples of fractons are given by certain symmetric rank-two tensor gauge theories \cite{Pretko:2016kxt, Bulmash:2018lid, Ma:2018nhd, Bulmash:2018knk}.  This paper will present examples of continuum fracton theories involving higher-rank fields that are neither symmetric nor antisymmetric tensors. One such example, for instance, will come from a theory defined on a cubic lattice with U(1)-valued fields on plaquettes; assuming that this theory is described by (anti)symmetric tensor gauge fields in the continuum limit will prove to be inconsistent at the level of equations of motion. This does not signal a sickness of these generalized fracton theories, but merely indicates that trying to recast fracton theories in terms of rotationally-covariant variables is not always the right way to go. This conclusion extends an idea that germinated in \cite{Ma:2018nhd}.
  \item The mathematical concepts used to systematically construct new  fracton theories in this paper are themselves rather novel, and they may prove to lead to a new cohomology theory with attendant new insights in algebraic topology. The new ideas are all centered on the notion of a generalized boundary operator $\del_2$. Roughly speaking, $\del_2$ acts on surfaces and returns their corners. Gauge theories based on the corresponding coboundary operator are the main subject of this paper.
\end{enumerate}

\subsection{The $\Ff_p$ classification of gauge theories}

Another way to motivate the present work is to cast it as a step forward in the exploration of the space of possible and physically interesting gauge constraints. Indeed, developments quite unassociated with fractons have recently led to an increased interest in generalizing the familiar concept of gauge theory featured in the Standard Model or in the standard textbooks.

On the one hand, higher-form gauge theories \cite{Wegner:1984qt} were lately placed into the context of studying higher-form symmetries in \cite{Gaiotto:2014kfa} and its follow-up papers: the main idea in this body of work is that gauge fields or symmetry currents need not be one-forms, but instead can be $k$-forms as well. A $k$-form gauge theory has a $(k - 1)$-form gauge parameter (or a $(k - 1)$-form charge density). On a lattice, such a theory can be realized by a field living on $k$-simplices (or, more generally, $k$-cells), with local gauge transformations acting on all $k$-cells that share the same $(k - 1)$-cell.

On the other hand, while ordinary gauge theories with modified gauge constraints have been known for a long time --- for instance in the guise of Chern-Simons theories or the Witten effect \cite{Witten:1979ey} --- it was found only relatively recently that such theories can be defined on par with ordinary gauge theories directly on the lattice \cite{Chen:2017fvr, Chen:2018nog}, with gauge constraints now incorporating both electric and magnetic fields. These are natural examples of theories with \emph{anomalous} higher-form symmetries.

The aim of this paper is to point out that fractons are found in a \emph{further} generalization along the first of these two paths. Given a theory of rank-$k$ fields, it is in fact possible to formulate a sequence of gauge constraints in which the gauge parameter is a rank-$(k - p)$ field. Each choice of $p$ gives a completely different class of gauge theories. These classes will be labeled $\Ff_p$, with the fraktur letter $\Ff$ chosen to noncommittally refer to fractons.

Gauge theories in the $\Ff_1$ class are standard gauge theories; $\Ff_1$ gauge fields with rank $k$ are the $k$-form gauge fields mentioned above. Gauge theories in the $\Ff_2$ class are the subject of this work. Theories in classes $\Ff_p$ for $p > 2$ will only be briefly referred to in the Conclusion; they are essentially trivial in $3+1$ or fewer dimensions. Generalizations of $\Ff_p$ theories in the second of the above directions --- to include flux attachment in their Gauss laws --- will be covered in future work, though related ideas have already appeared in \cite{Pretko:2017xar}.

All $\Ff_p$ theories with $p > 1$ are generically nonrelativistic. This can be seen heuristically by trying to construct the requisite path integrals. The temporal gauge fields must have rank $(k - p + 1)$, as they act as Lagrange multipliers that enforce gauge constraints. Meanwhile, spatial gauge fields have rank $k$ for any $p$. Without introducing new fields, it appears impossible to get a relativistic theory at $p > 1$, as the temporal and spatial gauge fields are of different rank.

\subsection{Overview of this paper}

This paper is self-contained, and a familiarity with fracton literature is not assumed. The analysis is almost exclusively carried out in the Hamiltonian framework,  with continuous  time and a closed 3d spatial lattice (a three-torus with size $L_x \times L_y \times L_z$ can be kept in mind at all times). It may be useful to be familiar with basic notions of Hamiltonian lattice gauge theory, for which the standard review is \cite{Kogut:1979wt}. The lattice spacing will be set to unity throughout.

Section \ref{sec Xcube} introduces the X-cube model, the simplest (though not the first) example of a theory showing fracton physics. The version of the X-cube model studied here is a strange $\Z_2$ gauge theory on a cubic lattice: its lines of electric flux can either move straight or split into two lines perpendicular to each other and the original line. The X-cube model is typically taken to refer to one specific Hamiltonian, but here the name will be used slightly more broadly to refer to all theories that satisfy these strange new gauge constraints. The focus will primarily be on the \emph{kinematics} of these theories: all gauge-invariant degrees of freedom and operators will be accounted for, kinematic symmetries will be analyzed before even introducing any Hamiltonian, and the discussion will be phrased in such a way to easily generalize in other sections.

Section \ref{sec dual Xcube} constructs the Kramers-Wannier dual of the X-cube model. While the construction is relatively straightforward, extra care will be devoted to getting the constraints and the topological degrees of freedom right on both sides of the duality. This is an exact duality, meaning that all operators and states map, regardless of the particular choice of the Hamiltonian for the X-cube degrees of freedom.

Section \ref{sec F2} explores the structure of the dual X-cube model and its component theories in great detail. Unlike the usual presentation of the X-cube model as a theory of rank-one $\Z_2$ fields, its dual presentation has a natural generalization to arbitrary lattices and to all gauge groups: a rank-two gauge theory of type $\Ff_2$ coupled to scalar matter. The case of continuous gauge groups will prove particularly interesting, as there it is possible to restrict the analysis to small fluctuations of gauge and matter fields, getting continuum Lagrangians that can be compared to the ones in the fracton literature. The results announced in this Introduction will all be presented there.

Finally, ideas for future work will be mentioned in Section \ref{sec outlook}. A handy summary of the most important models studied in this paper will also be given there.

\section{The X-cube model} \label{sec Xcube}

\subsection{Kinematics}

The X-cube model is defined on a spatial 3d cubic lattice $\Mbb$ with periodic boundary conditions. Each link on the lattice hosts a two-dimensional Hilbert space, just like in the setup of the toric code \cite{Kitaev:1997wr}. The operator algebra of the system, prior to imposing any gauge constraints, is the direct product of operator algebras on links $\ell$. The algebra on each link is generated by a position operator $Z_\ell$ and a kinetic operator $X_\ell$, which are just the usual Pauli matrices.

An \emph{ordinary} $\Z_2$ gauge theory is obtained by imposing the local constraint
\bel{\label{ord G}
  G_v \equiv \prod_{\ell \supset v} X_\ell = \1
}
at each vertex $v$. The kinetic operators $X_\ell$ can be written as exponentials $\e^{\i \pi E_\ell}$ of $\Z_2$ electric fields $E_\ell$, and then the gauge constraints become $\sum_{\ell \supset v} E_\ell = 0$, which is the lattice version of the more familiar U(1) continuum Gauss law $\bfnabla \b E = 0$. Gauge-invariant operators are those that commute with the Gauss operators $G_v$. The gauge-invariant algebra is generated by electric operators $X_\ell$, Wilson loops $W_f \equiv \prod_{\ell \subset f} Z_\ell$ on plaquettes (faces) $f$ of the lattice, and by one Wilson loop $W_l \equiv \prod_{\ell \in l} Z_\ell$ from each nontrivial element $l$ of the cellular homology group $H_1(\Mbb, \Z_2)$. These operators all act within the gauge-invariant Hilbert space, which is spanned by all possible \emph{closed} loops of electric flux in the eigenbasis of the $X_\ell$ operators.

The version of the X-cube model studied here is a gauge theory, but it is \emph{not} ordinary. It has three constraints per vertex:
\bel{\label{Xcube G}
  G_v^{i} \equiv \prod_{\substack{\ell \supset v\\ \ell \perp i}} X_\ell = \1, \quad i \in \{x, y, z\}.
}
See Fig.\ \ref{fig Xcube} for an illustration. The gauge-invariant algebra is generated by the electric operators $X_\ell$, Wilson lines $W_s$ along all \emph{straight} one-cycles $s$, and Wilson cubes\footnote{The X-cube model is named after these operators --- in the original paper, $X_\ell$ denoted position operators, and hence $W_c$ was a product of 12 $X$ operators along a cube. The theory under consideration in this paper could have been more properly called the Z-cube model.}
\bel{
  W_c \equiv \prod_{\ell \in c} Z_\ell.
}
The gauge-invariant Hilbert space is spanned by cubes and closed straight lines of electric flux. The key feature of the model is that electric flux lines cannot bend: they must either go straight until they fully wind around the spatial torus, or one line can split into two lines that are perpendicular to each other and to the original one.

% commands from http://www.texample.net/tikz/examples/3d-graph-model/

\newcommand{\xytransf}[2]
{ % first 4 params = transf matrix that changes the coordinate system
  % 5th param = origin of the transformation
    \pgftransformcm{1}{0}{0.4}{0.4}{\pgfpoint{#1cm}{#2cm}}
}

\newcommand{\yztransf}[2]
{
   \pgftransformcm{0.4}{0.4}{0}{1}{\pgfpoint{#1cm}{#2cm}}
}

\newcommand{\gridThreeD}[5]
{
    \begin{scope}
        \xytransf{#1}{#2}; %first realign TikZ's coords
        \draw [#3,step= 1cm] grid (#4,#5); %draw a "normal" grid
    \end{scope}
}

\begin{figure}
\begin{center}
\begin{tikzpicture}[scale = 2]

\draw[step = 1, dotted] (0, 0) grid (2, 2);
\draw[step = 1, dotted, xshift=0.4cm, yshift=0.4cm] (0.0, 0.0) grid (2, 2);
\draw[step = 1, dotted, xshift=0.8cm, yshift=0.8cm] (0.0, 0.0) grid (2, 2);
\gridThreeD{0}{0}{dotted}22;
\gridThreeD{0}{1}{dotted}22;
\gridThreeD{0}{2}{dotted}22;

\begin{scope}
  \xytransf02;

  \draw[blue, thick] (0.05, 1) -- (1.95, 1);
  \draw[blue, thick] (1, 0.05) -- (1, 1.95);

  \draw[blue] (1, 1) node[above left] {\contour{white}{$G^z_{v_3}$}};
\end{scope}

\begin{scope}
  \yztransf20;

  \draw[red, thick] (1, 0.05) -- (1, 1.95);
  \draw[red, thick] (0.05, 1) -- (1.95, 1);

  \draw[red] (1, 1) node[below right] {\contour{white}{$G^x_{v_1}$}};
\end{scope}

\begin{scope}
  \yztransf10;

  \draw[olive, thick] (1, 0.05) -- (1, 1.95);
  \draw[olive] (1, 1) node[below left] {\contour{white}{$G^y_{v_2}$}};
\end{scope}

\begin{scope}
  \xytransf01;

  \draw[olive, thick] (0.05, 1) -- (1.95, 1);
\end{scope}

\draw[->, thick] (0, 0) -- (0, 0.4);
\draw (0.4, 0) node[above right] {\contour{white}{$x$}};

\draw[->, thick] (0, 0) -- (0.4, 0);
\draw (0, 0.4) node[below left] {\contour{white}{$z$}};
\begin{scope}
  \yztransf00;

  \draw[->, thick] (0, 0) -- (0.6, 0);
  \draw (0.6, 0) node[above right] {\contour{white}{$y$}};
\end{scope}

\draw[step = 1, dotted] (4, 0) grid (6, 2);
\draw[step = 1, dotted, xshift=0.4cm, yshift=0.4cm] (4, 0.0) grid (6, 2);
\draw[step = 1, dotted, xshift=0.8cm, yshift=0.8cm] (4, 0.0) grid (6, 2);
\gridThreeD{4}{0}{dotted}22;
\gridThreeD{4}{1}{dotted}22;
\gridThreeD{4}{2}{dotted}22;

\begin{scope}
  \yztransf40;

  \draw[purple, thick] (0,0) grid (1, 1);
  \draw[purple, thick] (2, -0.2) -- (2, 2.2);
  \filldraw[draw = white, fill=white] (2, 1.6) circle [radius = 1pt];
  \filldraw[draw = white, fill=white] (2, 0.6) circle [radius = 1pt];
  \filldraw[draw = white, fill=white] (2, 0.2) circle [radius = 1pt];
\end{scope}

\begin{scope}
  \yztransf50;

  \draw[purple, thick] (0,0) grid (1, 1);
\end{scope}

\begin{scope}
  \xytransf40;

  \draw[purple, thick] (0,0) grid (1, 1);
\end{scope}

\begin{scope}
  \xytransf41;

  \draw[purple, thick] (0,0) grid (1, 1);
\end{scope}

\begin{scope}
  \xytransf42;

  \draw[purple, thick] (-0.2, 1) -- (2, 1);
  \draw[purple, thick] (2, 1) -- (2, 2.2);
\end{scope}

\begin{scope}
  \yztransf60;

  \draw[purple, thick] (1, 2) -- (1, -0.2);
\end{scope}

\end{tikzpicture}
\end{center}
\caption{\small \textit{Left}, A small part of a cubic lattice, with three Gauss operators \eqref{Xcube G} on three different sites $v_1$, $v_2$, $v_3$ indicated in red, green, and blue. Each Gauss operator is a product of electric operators $X_\ell$ on links of the same color. Links used to construct each $G_v^i$ emanate from $v$ and are perpendicular to direction $i$. \textit{Right}, the configuration of electric flux lines in an example gauge-invariant state: flux lines either go straight or split into two lines perpendicular to each other and to the original line. A ``glueball'' of electric flux lives on the cube in the lower left corner.
}
\label{fig Xcube}
\end{figure}
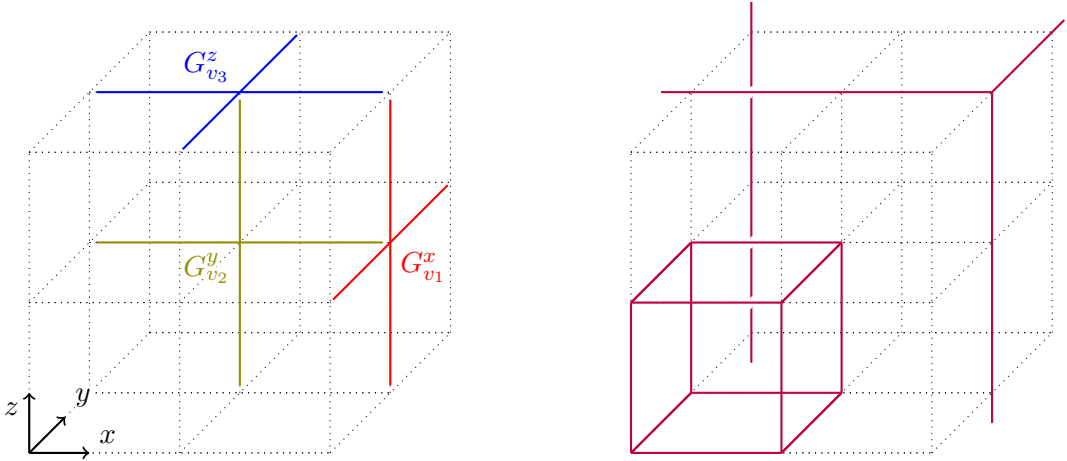

Local excitations in this theory can be viewed in two natural ways. In lattice gauge theory jargon, \emph{glueballs} are excitations made up of closed lines of electric flux, and here they live on cubes and not on plaquettes, as they do in the ordinary gauge theory. They are created and destroyed by $W_c$ operators, but they cannot be individually measured because $X_\ell$, which measures electric flux, always touches four cubes at once. The implication of this is clearer from the second point of view on excitations, described below.

An approach more germane to condensed matter is to view local excitations as eigenstates of magnetic operators $W_c$. These are called \emph{fractons} in this context, and they live on cubes just like glueballs. Because each link touches four cubes, fractons can only be created in sets of four.\footnote{This situation is different in the presence of boundaries, of course, but the analysis here will be restricted to a spatial three-torus.} There is thus no sense in which a single fracton is a stable propagating particle, but it is still possible to interpret a pair  of adjacent fractons (i.e.~a \emph{dipole}) as being propagated by the $X_\ell$'s along a direction perpendicular to the dipole's orientation. The number of these dipoles (mod 2) is conserved  \emph{separately} in each plane because of the restricted motion afforded to any dipole.

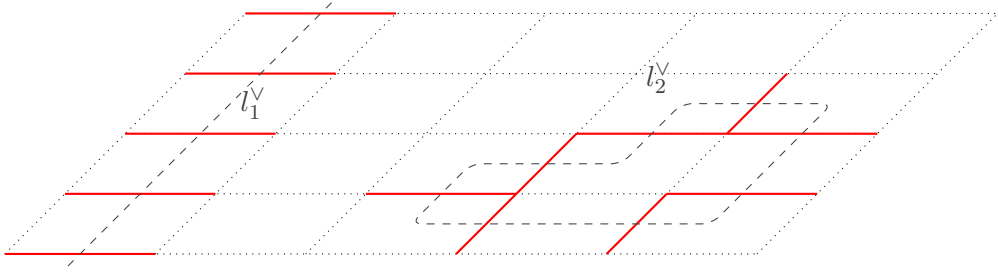
\begin{figure}
\begin{center}
\begin{tikzpicture}[scale = 2]

\gridThreeD{0}{0}{dotted}54;

\begin{scope}
  \xytransf00;
  \foreach \x in {0, ..., 4} {
    \draw[red, thick] (0, \x) -- (1, \x);
    %\draw[red] (0.5, \x) node[above left] {\contour{white}{$X$}};
  };
  \draw[darkgray, dashed] (0.5, -0.2) -- (0.5, 4.2);
  \draw[darkgray] (0.5, 2.5) node[right] {\contour{white}{$l^\vee_1$}};

  \draw[red, thick] (3, 1) -- (2, 1);
  \draw[red, thick] (3, 1) -- (3, 0);
  \draw[red, thick] (3, 1) -- (3, 2);

  \draw[red, thick] (4, 1) -- (4, 0);
  \draw[red, thick] (4, 1) -- (5, 1);

  \draw[red, thick] (4, 2) -- (4, 3);
  \draw[red, thick] (4, 2) -- (5, 2);
  \draw[red, thick] (4, 2) -- (3, 2);

  \draw[dashed, rounded corners, darkgray] (3.5, 0.5) -- (2.5, 0.5) -- (2.5, 1.5) -- (3.5, 1.5) -- (3.5, 2.5) -- (4.5, 2.5) -- (4.5, 0.5) -- (3.5, 0.5);

  \draw[darkgray] (3.5, 2.5) node[above left] {\contour{white}{$l_2^\vee$}};
\end{scope}

\end{tikzpicture}
\end{center}
\caption{\small Two examples of two-form electric symmetry operators in a single plane $p$ of the cubic lattice. The operators $U_{l^\vee_1}$ and $U_{l^\vee_2}$ from \eqref{def Ul} are defined in terms of closed lines $l^\vee_1$ and $l^\vee_2$ on the square lattice dual to the slice $p$: the operators are given by products of all $X_\ell$'s whose links pierce the lines $l^\vee$. The line $l^\vee_1$ wraps around the torus and the associated generator is nontrivial: the line can be smoothly deformed within $p$ but cannot be shrunk to a point. On the other hand, $U_{l_2^\vee}$ is always equal to the identity because it can be represented as a product of Gauss operators $G^z$.
}
\label{fig Xcube symmetries}
\end{figure}

Just like any gauge theory, the X-cube model contains some kinematic symmetries, i.e.~symmetries whose generators commute with all local gauge-invariant operators. In ordinary gauge theories, these are called electric one-form symmetries \cite{Gaiotto:2014kfa}. Their generators are topological, being products of electric operators $X_\ell$ over closed codimension-one manifolds, with homologically equivalent manifolds giving rise to identical symmetry generators due to the constraints \eqref{ord G}. The theory at hand, however, has \emph{electric two-form symmetries} that are topological in a limited sense only. The generators of these symmetries live on codimension-two manifolds (i.e.~lines) and are given by
\bel{\label{def Ul}
  U_{l^\vee} \equiv \prod_{\ell \subset l^\vee} X_\ell,
}
where $l^\vee$ is a one-cycle on the dual lattice that completely lies in one plane; a link $\ell$ belongs to the line $l^\vee$ if the link dual to $\ell$ is in $l^\vee$, with the lattice duality performed \emph{only} in the speficied plane (see Fig.~\ref{fig Xcube symmetries}). The lines $l^\vee$ may be deformed with impunity only by multplying $U_{l^\vee}$ by Gauss operators $G^{i}_v$, with $i$ being the one direction perpendicular to $p$. In particular, this means that for every plane that intersects the spatial torus, there is an \emph{independent} symmetry generator.

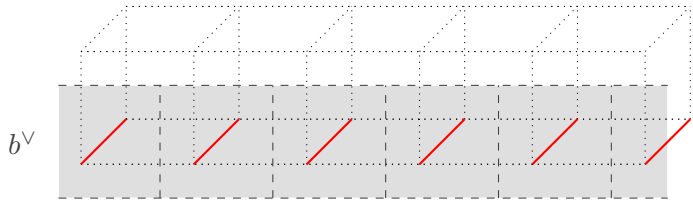
\begin{figure}[t]
\begin{center}
\begin{tikzpicture}[scale = 1.5]

\filldraw[fill = lightgray!50, draw = white] (-0.2, -0.3) rectangle (5.2, 0.7);

\gridThreeD{0}{0}{dotted}51;
\gridThreeD{0}{1}{dotted}51;
\draw[step = 1, dotted] (0, 0) grid (5, 1);
\draw[step = 1, dotted, xshift=0.4cm, yshift=0.4cm] (0.0, 0.0) grid (5, 1);

\begin{scope}
  \xytransf00;
  \foreach \x in {0, ..., 5}
    \draw[red, thick] (\x, 0) -- (\x, 1);
\end{scope}

\draw[darkgray, dashed] (-0.2, -0.3) -- (5.2, -0.3);
\draw[darkgray, dashed] (-0.2, 0.7) -- (5.2, 0.7);
\foreach \x in {0, ..., 4}
  \draw[darkgray, dashed] (\x + 0.7, -0.3) -- (\x + 0.7, 0.7);

\draw[darkgray] (-0.3, 0) node[above left] {$b^\vee$};
\end{tikzpicture}
\end{center}
\caption{\small The shaded area is a two-chain $b^\vee$ on the dual lattice, called a belt. The red links it corresponds to on the original lattice host the operators that make up electric two-form symmetry generators $U_{b^\vee}$.
}
\label{fig belts}
\end{figure}

It is a bit cumbersome to work with dual lattices in each plane separately. Instead, the independent generators of the two-form symmetries in the X-cube model will be taken to be
\bel{\label{2form symms}
  U_{b^\vee} \equiv \prod_{\ell \subset b^\vee} X_\ell,
}
where $b^\vee$ is a closed \emph{belt} of dual plaquettes (in the full, 3d sense of duality; see Fig.~\ref{fig belts}). The links $\ell \in \Mbb$ that belong to $b^\vee$ are all parallel to each other, they lie in the same plane, and they wrap a noncontractible cycle of the three-torus.

It may be illuminating to count the degrees of freedom in this theory (see also \cite{Shirley:2017suz}). Let the three-torus $\Mbb$ have $L_x \times L_y \times L_z$ sites. Before any gauge constraints were imposed, the system had $3 L_x L_y L_z$ $\Z_2$ degrees of freedom --- one per link. The number of independent constraints is $2L_x L_y L_z - (L_x + L_y + L_z) + 1$ --- two per site because $G^{(z)}_v = G^{(x)}_v G^{(y)}_v$, minus one relation $\prod_{v \subset p} G_v^{i} = \1$ per plane $p$, with $i \perp p$, plus one because the number of relations is overcounted: taking the product of $G_v$'s along all planes perpendicular to the $x$ and $y$ directions is the same as taking the product of $G_v$'s along just the planes parallel to the $z$ direction. Thus the total number of gauge-invariant degrees of freedom is $L_x L_y L_z + L_x + L_y + L_z - 1$.

The number of glueball or fracton degrees of freedom included in the above count corresponds to the number of linearly independent Wilson cubes, which is $L_x L_y L_z - (L_x + L_y + L_z) + 2$ --- one per cube, minus one relation per plane, plus two because of the overcounting of relations (multiplying all Wilson cubes in planes parallel to any direction gives the same result). The remaining $2(L_x + L_y + L_z) - 3$ degrees of freedom are the linearly independent straight lines of electric flux, modulo those lines that can be built out of glueballs. Independently counting these also gives $L_x L_y + L_y L_z + L_z L_x - (L_x - 1)(L_y - 1) - (L_y - 1)(L_z - 1) - (L_z - 1)(L_x - 1) = 2(L_x + L_y + L_z) - 3$.

%The number of local excitations (whether fractons or glueballs) is $L_x L_y L_z - (L_x + L_y + L_z - 3)$, because the product of Wilson cubes $W_c$ along any plane $p$ is necessarily trivial.

\subsection{Dynamics}

The discussion so far was completely kinematical. What kinds of natural Hamiltonians can be defined for the X-cube degrees of freedom? The one most commonly discussed in the literature describes the topological phase in which all fractons are gapped out,
\bel{\label{H0}
  H_0 = \sum_c W_c.
}
In this theory only the straight lines of electric flux remain at low energies. The ground state degeneracy is
\bel{
  \trm{GSD} = 2^{2(L_x + L_y + L_z) - 3}.
}
One remarkable property of this model is precisely this result: the number of ground states in the topological phase increases exponentially with the linear size of the system.

More generally, a natural Hamiltonian for the X-cube model takes the Kogut-Susskind form \cite{Kogut:1974ag}
\bel{\label{Hg}
  H_g = -\frac{g^2}{\pi^2} \sum_\ell X_\ell - \frac1{g^2} \sum_c W_c,
}
with the topological phase obtained at $g = 0$. At $g \rar \infty$, the theory is in the confined phase with a unique ground state. Thus there must exist a strong-weak coupling phase transition --- presumably a first-order one.

\section{The dual X-cube model}\label{sec dual Xcube}

The dual description of the X-cube model (in the sense of Kramers and Wannier) has been presented, with various degrees of completeness, in \cite{Vijay:2016phm, Williamson:2016jiq, Ma:2018nhd, Weinstein:2018xil}.  It is simplest to start this discussion with the local mapping of operators
\algns{\label{dual local}
  W_c &= X_c^\vee, \\
  X_\ell &= \prod_{c \supset \ell} Z_c^\vee,
}
which maps the X-cube operators $W_c$ and $X_\ell$ into operators that act on Ising spins on the dual lattice $\Mbb^\vee$. Checks ($^\vee$) denote operators in this dual spin system. Each cube/plaquette/link/vertex of the original lattice corresponds uniquely to a vertex/link/plaquette/cube of the dual lattice, so labels for lattice components like $c$, $v$, and $\ell$ can be used to unambiguously label the dual elements too: a checked variable (e.g.~$X_c^\vee$) should be understood to live on the dual vertex $c$ in $\Mbb^\vee$.

A local duality like \eqref{dual local} does not map all degrees of freedom to each other. The Wilson lines $W_s$ do not have duals under this mapping. This is reflected at the level of Hilbert spaces by the fact that the electric two-form symmetry \eqref{2form symms} must be in a specific superselection sector, $U_{b^\vee} = \1$ (for all dual belts $b^\vee$), for this duality to be globally consistent. Similarly, the algebra of the dual spin system generated by $X_c^\vee$ and $\prod_{c \supset \ell} Z_c^\vee$ has a center generated by
\bel{\label{def Q vee}
  Q_{p^\vee}^\vee \equiv \prod_{c \subset p^\vee} X_c^\vee.
}
These operators are defined for each plane $p^\vee$ of the dual lattice (compare this to operators $U_{l^\vee}$ defined on closed lines in the dual lattice). For the duality \eqref{dual local} to be consistent, the spin system must be in the $Q_{p^\vee}^\vee = \1$ sector for each dual plane. Thus this local operator duality can be written as
\boxedAlgns{\label{dual local scheme}
  \frac{\trm{X-cube\ model}}{\{U_{b^\vee}\}} = \frac{\trm{Ising\ spins}^\vee}{\{Q^\vee_{p^\vee}\}}.
}

Note that the usual 3d Kramers-Wannier paradigm dualizes a spin system to a two-form gauge theory, and an ordinary gauge theory to another gauge theory (this is S-duality, or the electric-magnetic duality). The removal of $Z^\vee$ bilinears  changes this paradigm completely: with only four-operator combinations of $Z^\vee$, the spin system ends up being mapped to an unusual gauge theory instead.

To find the dual of the full X-cube model, the local duality above should first be \emph{twisted} by introducing classical $\Z_2$ gauge fields on dual plaquettes $\eta_\ell^\vee$. (This procedure is explained in great detail for ordinary dualities of Kramers-Wannier and Jordan-Wigner type in \cite{Radicevic:2018okd}.) The twisted duality is
\algns{\label{dual twisted}
  W_c &= X_c^\vee, \\
  X_\ell &= \eta_\ell^\vee \prod_{c \supset \ell} Z_c^\vee.
}
It differs from the original duality \eqref{dual local} because now the two-form symmetries are no longer in their singlet sectors,
\bel{
  U_{b^\vee} = \prod_{\ell \supset b^\vee} \eta^\vee_\ell.
}
Thus, by judiciously choosing the background gauge fields $\eta^\vee$, different sectors of the X-cube model can individually be mapped to a spin system. These background fields can be assumed to satisfy $\prod_{\ell \supset l^\vee} \eta^\vee_\ell = 1$ for a homologically trivial $l^\vee$; if they do not, the duality is inconsistent unless the X-cube model is coupled to classical background charges.

\begin{figure}
\begin{center}
\begin{tikzpicture}[scale = 1.5]

\draw[step = 1, dotted] (0, 0) grid (2, 2);
\draw[step = 1, dotted, xshift=0.4cm, yshift=0.4cm] (0.0, 0.0) grid (2, 2);
\draw[step = 1, dotted, xshift=0.8cm, yshift=0.8cm] (0.0, 0.0) grid (2, 2);
\gridThreeD{0}{0}{dotted}22;
\gridThreeD{0}{1}{dotted}22;
\gridThreeD{0}{2}{dotted}22;

\begin{scope}
  \yztransf00;

  \draw[blue, thick] (0,0) grid (1, 1);
\end{scope}

\begin{scope}
  \yztransf10;

  \draw[blue, thick] (0,0) grid (1, 1);
\end{scope}

\begin{scope}
  \xytransf00;

  \draw[blue, thick] (0,0) grid (1, 1);
\end{scope}

\begin{scope}
  \xytransf01;

  \draw[blue, thick] (0,0) grid (1, 1);
\end{scope}

\begin{scope}
  \xytransf01;

  \draw[red, thick] (1.1, 1) -- (1.9, 1);
  %\draw[purple, thick] (2, 1) -- (2, 2.2);
\end{scope}

\draw[blue] (1, 1.2) node[above left] {\contour{white}{$W_c$}};
\draw[red] (2, 1.4) node[above] {\contour{white}{$X_\ell$}};

\begin{scope}
  \yztransf{5.55}{0.7};

  \filldraw[red, fill = red!20, thick] (0, 0) rectangle (1, 1);
  \draw[red, thick] (0, 0) node {$\bullet$};
  \draw[red, thick] (1, 0) node {$\bullet$};
  \draw[red, thick] (0, 1) node {$\bullet$};
  \draw[red, thick] (1, 1) node {$\bullet$};
\end{scope}

\draw[step = 1, dotted, gray] (4, 0) grid (6, 2);
\draw[step = 1, dotted, gray, xshift=0.4cm, yshift=0.4cm] (4, 0.0) grid (6, 2);
\draw[step = 1, dotted, gray, xshift=0.8cm, yshift=0.8cm] (4, 0.0) grid (6, 2);
\gridThreeD{4}{0}{dotted, gray}22;
\gridThreeD{4}{1}{dotted, gray}22;
\gridThreeD{4}{2}{dotted, gray}22;

\draw[blue, thick] (4.7, 0.7) node {$\bullet$};
\draw[blue] (4.7, 0.7) node[above left] {\contour{white}{$X_c^\vee$}};

\draw[red] (5.55, 0.7) node[below] {\contour{white}{$Z_{c_1}^\vee$}};
\draw[red] (5.55, 1.7) node[left] {\contour{white}{$Z_{c_2}^\vee$}};
\draw[red] (6, 2.1) node[above] {\contour{white}{$Z_{c_3}^\vee$}};
\draw[red] (6, 1.1) node[right] {\contour{white}{$Z_{c_4}^\vee$}};

\draw[red] (5.75, 1.45) node {\contour{white}{$Z_\ell^\vee$}};

\end{tikzpicture}
\end{center}
\caption{\small \emph{Left}, the local operators $W_c$ and $X_\ell$ of the X-cube model. \emph{Right}, their dual operators, as per eq.~\eqref{dual global}. The cube operator $W_c$, shown in blue, maps to a matter momentum operator $X_c^\vee$ that lives on the site dual to cube $c$. The electric field operator $X_\ell$, shown in red, maps to the product of five operators: a gauge field position operator $Z_\ell^\vee$ on the plaquette dual to link $\ell$, and four matter position operators $Z_{c_i}^\vee$ living on sites dual to the cubes $c_i$ that share the link $\ell$.
}
\label{fig duality}
\end{figure}
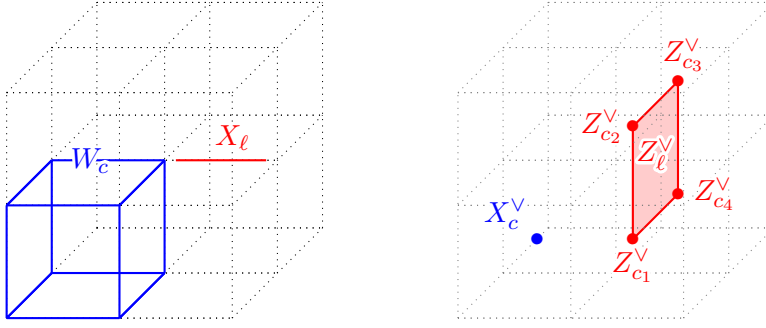

The final step is to make the background fields $\eta^\vee$ dynamical. This is done by replacing the classical fields with appropriate position operators,
\bel{
  \eta^\vee_\ell \mapsto Z^\vee_\ell,
}
and by introducing their conjugate operators $X^\vee_\ell$ (the $\Z_2$ rank-two electric fields). It is now simple to check that the following global duality is consistent (see Fig.~\ref{fig duality} for an illustration of how local operators map):
\gathl{\label{dual global}
  W_c = X_c^\vee, \qquad
  X_\ell = Z_\ell^\vee \prod_{c \supset \ell} Z^\vee_c,\\
  W_{s}  = \prod_{\ell \subset s} X^\vee_\ell,
}
supplemented by a gauge constraint
\bel{\label{dual G}
  G_c^\vee \equiv X_c^\vee \prod_{\ell \supset c} X_\ell^\vee = \1,
}
and a flatness condition for every contractible dual belt
\bel{\label{flatness}
  \prod_{\ell \subset b^\vee} Z_\ell^\vee = \1.
}

This duality can be written as
\boxedAlgns{\label{dual global scheme}
  \trm{X-cube\ model} = \frac{\trm{Ising\ spins}^\vee}{\{Q^\vee_{p^\vee}\}} \times \Z_2 \trm{\ rank-two\ topological\ $\Ff_2$\ gauge\ theory}.
}
It may be helpful to unpack this nomenclature. The Ising spins live on dual sites, and their operators are $X_c^\vee$ and $Z_c^\vee$. The Ising degrees of freedom are in the singlet sector of each of the $\Z_2$ symmetries generated by $Q^\vee_{p^\vee}$ from \eqref{def Q vee}. The $\Z_2$ gauge fields live on dual plaquettes (hence they are rank-two) and their operators are $X_\ell^\vee$ and $Z_\ell^\vee$. They are called $\Ff_2$ gauge fields because their gauge charges live on objects two ranks below theirs (unlike in ordinary gauge theories where gauge fields are $k$-forms and charged fields are $(k - 1)$-forms). The gauge-invariant and flatness-preserving combinations of $X_\ell^\vee$ and $Z_\ell^\vee$ are precisely the operators that appear on the r.h.s.~of \eqref{dual global}. Because the gauge fields must be locally flat, the gauge theory is called topological.

%The dual of the X-cube model is an unusual theory built out of usual components --- Ising spins and $\Z_2$ 2-form gauge fields. The \emph{only} unusual feature is their coupling. In ordinary theories, $p$-form degrees of freedom are naturally coupled to $(p+1)$-form ones. Generalizing this rule to couplings of $p$-form degrees of freedom to $(p + k)$-form ones leads to $\frak F_k$ gauge theories.

The archetypical theory \eqref{Hg} dualizes into
\bel{
  H_g = \frac{g^2}{\pi^2} \sum_\ell Z_\ell^\vee \prod_{c \supset \ell} Z^\vee_c - \frac1{g^2} \sum_c X_c^\vee.
}
It is now evident that in the $g \rar 0$ limit, the two fields decouple, with the Ising spins being in the paramagnetic phase and the rank-two gauge fields being completely degenerate.\footnote{The dual X-cube model always has topological (i.e.~flat) rank-two gauge fields that satisfy eq.~\eqref{flatness}. This condition is not energetically imposed, and no Hamiltonian dual to X-cube degrees of freedom can have individual $X_\ell^\vee$ terms that would give rise to nontrivial curvature.} This is the dual interpretation of the fracton topological order. In the opposite limit, $g \rar \infty$, the ground state is unique and corresponds to a breaking up of ``mesons'' --- bound states of quadruplets of $\Z_2$ excitations --- with the resulting deconfined ``plasma'' destroying the topological order.

\section{Gauge theories of type $\Ff_2$} \label{sec F2}

A key property of the dual X-cube model is that it can be naturally defined on an arbitrary 3d lattice $\Mbb$. The Gauss law \eqref{dual G} does not depend on the lattice being cubic, and the flatness condition is naturally generalized to the requirement that all local magnetic operators must be equal to the identity. The goal of this Section is to develop a basic description of such theories on arbitrary lattices.

\subsection{Generalized boundary operators}

A sufficiently versatile description of these new gauge theories requires some novel mathematical definitions. To start, here is a brief review of notation used for more familiar mathematical structures. Given a $d$-dimensional lattice $\Mbb$, it is natural to define vector spaces of $k$-chains $C_k(\Mbb, \Z_K)$, with $\Z_K$ being the scalar field over which the vectors are defined. For purposes of $\Z_K$ gauge theories, it is sufficient to study $C_k(\Mbb, \Z_K)$; here the focus will be on gauge groups $\Z_2$ and $\trm U(1) = \lim_{K \rar \infty} \Z_K$, with coefficients in $\Z_2$ and $\Z$, respectively. A boundary operator is a map $\del\!: C_k \mapsto C_{k - 1}$ that takes a $k$-chain and returns some natural linear combination of its consitutent $(k - 1)$-chains. In the case of $C_k(\Mbb, \Z)$, the definition of $\del$ involves a choice of orientation. In other words, for any elementary $k$-chain (a.k.a.~a $k$-cell) $\omega^{(k)} \in C_k(\Mbb, \Z)$, its boundary is defined to be the linear combination of $(k - 1)$-cells
\bel{\label{def del}
  \del \omega^{(k)} \equiv \sum_i (-1)^{\sigma_i} \omega_i^{(k - 1)} \quad \trm{for}\quad \omega_i^{(k - 1)} \subset \omega^{(k)},
}
and the choice $\sigma_i \in \{0, 1\}$ for every $k$-cell  defines an orientation of $\Mbb$. It is standard to choose orientations such that
\bel{
  \del^2 = 0,
}
and if this is not possible, the lattice $\Mbb$ is said to be unorientable.

Another natural operator on the space of all possible chains is $\del_{-1}\!: C_k \mapsto C_{k + 1}$, which acts as
\bel{\label{def del -1}
  \del_{-1} \omega^{(k)} \equiv \sum_i (-1)^{\tau_i} \omega_i^{(k + 1)},
}
with the sum running over all $(k + 1)$-cells $\omega_i^{(k + 1)}$ that have $\omega^{(k)}$ in their boundary. It is closely related to the coboundary operator $\delta$, which acts on sets of cochains (linear functions on chains) as $\delta\!: C^k \mapsto C^{k + 1}$ --- just like the exterior derivative acts on $k$-forms in differential geometry. Further, the coefficients $\tau_i$ are naturally induced from the $\sigma$'s that enter the definition of the boundary operator $\del$ in \eqref{def del}.

To illustrate the relation between $\del$, $\delta$, and $\del_{-1}$, consider the case of ordinary U(1) gauge theory on $\Mbb$. The Gauss operators are
\bel{
  G_v \equiv \prod_{\ell \in \del_{-1} v} X_\ell.
}
The above formula is a shorthand for\footnote{Note that $\subset$ is used only to describe that a $k$-cell belongs to a $k'$-cell (or a set of them) for $k' > k$. From here on, $\in$ is used to describe that a $k$-cell belongs to a $k$-chain from $C_k(\Mbb, \Z)$. Products over chains are defined as e.g.~
 \bel{
   \prod_{v \in \omega} \O_v = \prod_v \O_v^{n_v} \quad \trm{for}\quad \omega = \sum_v n_v v.
 }
 }
\bel{
  G_v = \prod_{\ell \supset v} X_{\ell}^{(-1)^{\tau_\ell(v)}}, \quad \trm{where} \quad \del_{-1} v = \sum_{\ell \supset v} (-1)^{\tau_\ell(v)} \ell.
}
Now consider an arbitrary zero-cochain $n: C_0 \mapsto \Z$. A generic gauge transformation is
\bel{
  G[n] \equiv \prod_v G_v^{n_v} = \prod_v \prod_{\ell \in \del_{-1}v} X_\ell^{n_v}.
}
It is possible  to reorganize this product into
\bel{
  G[n] = \prod_\ell \prod_{v \in \del \ell} X_\ell^{n_v} \equiv \prod_\ell X_\ell^{(\delta n)_\ell},
}
which is simply the statement that a gauge transformation changes the vector potential on each link by the derivative of a scalar field $n$ along that link. The coefficients $\sigma$ in \eqref{def del} and $\tau$ in \eqref{def del -1} --- as well as the ones that enter into the definition of $\delta$, which will not be needed here --- are related by the above equations for $G[n]$. These equations are in fact a version of Stokes' theorem on the lattice.

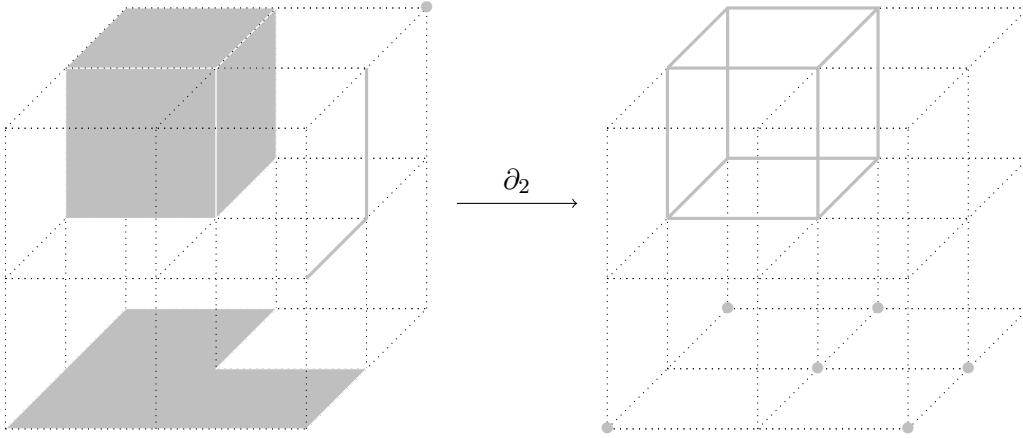
\begin{figure}
\begin{center}
\begin{tikzpicture}[scale = 2]

\gridThreeD{0}{1}{dotted}22;
\gridThreeD{0}{0}{dotted}22;
\draw[step = 1, dotted, xshift=0.4cm, yshift=0.4cm] (0.0, 0.0) grid (2, 2);
\draw[step = 1, dotted, xshift=0.8cm, yshift=0.8cm] (0.0, 0.0) grid (2, 2);

\begin{scope}
  \xytransf00;

  \filldraw[fill = lightgray, draw = white] (0, 0) -- (0, 2) -- (1, 2) -- (1, 1) -- (2, 1) -- (2, 0) -- (0, 0);
\end{scope}

\draw[dotted] (1.4, 0.8) -- (1.4, 0.4);

\filldraw[fill = lightgray, draw = white, xshift = 0.4cm, yshift = 0.4cm] (0, 1) -- (1, 1) -- (1, 2) -- (0, 2) -- (0,1);

\begin{scope}
  \yztransf10;

  \filldraw[fill = lightgray, draw = white] (1, 1) -- (2, 1) -- (2, 2) -- (1, 2) -- (1,1);
\end{scope}

\begin{scope}
  \xytransf02;

  \filldraw[fill = lightgray, draw = white] (0, 1) -- (1, 1) -- (1, 2) -- (0, 2) -- (0,1);
\end{scope}

\gridThreeD{0}{2}{dotted}22;
\draw[step = 1, dotted] (0, 0) grid (2, 2);

\draw[lightgray, very thick] (2, 1) -- (2.4, 1.4) -- (2.4, 2.4);

\draw[lightgray] (2.8, 2.8) node {$\bullet$};

\draw[->] (3, 1.5) -- (3.8, 1.5)
  node[midway, above]{$\del_2$};

\draw[step = 1, dotted] (4, 0) grid (6, 2);
\draw[step = 1, dotted, xshift=0.4cm, yshift=0.4cm] (4, 0.0) grid (6, 2);
\draw[step = 1, dotted, xshift=0.8cm, yshift=0.8cm] (4, 0.0) grid (6, 2);
\gridThreeD{4}{0}{dotted}22;
\gridThreeD{4}{1}{dotted}22;
\gridThreeD{4}{2}{dotted}22;

\begin{scope}
  \xytransf40;

  \draw[lightgray] (0, 0) node {$\bullet$};
  \draw[lightgray] (2, 0) node {$\bullet$};
  \draw[lightgray] (2, 1) node {$\bullet$};
  \draw[lightgray] (1, 1) node {$\bullet$};
  \draw[lightgray] (1, 2) node {$\bullet$};
  \draw[lightgray] (0, 2) node {$\bullet$};
\end{scope}

\begin{scope}
  \xytransf41;

  \draw[lightgray, very thick] (0, 1) -- (1, 1) -- (1, 2) -- (0, 2) -- (0, 1);
\end{scope}

\begin{scope}
  \xytransf42;

  \draw[lightgray, very thick] (0, 1) -- (1, 1) -- (1, 2) -- (0, 2) -- (0, 1);
\end{scope}

\draw[lightgray, very thick, xshift = 0.4cm, yshift = 0.4cm] (4, 1) -- (4, 2);
\draw[lightgray, very thick, xshift = 0.4cm, yshift = 0.4cm] (5, 1) -- (5, 2);
\draw[lightgray, very thick, xshift = 0.8cm, yshift = 0.8cm] (4, 1) -- (4, 2);
\draw[lightgray, very thick, xshift = 0.8cm, yshift = 0.8cm] (5, 1) -- (5, 2);
\end{tikzpicture}
\end{center}
\caption{\small The action of $\del_2$ on chains of all possible ranks in 3d. Two-chains (surfaces) are reduced to their corners, and three-chains (cubes) are reduced to their edges. One- and zero-chains are annihilated. If chains live in a vector space with integer coefficients, like $C_2(\Mbb, \Z)$, the resulting one- and zero-chains will typically have coefficients $\pm 1$ depending on the choice of orientation when defining $\del_2$.
}
\label{fig del2}
\end{figure}

So much for reviewing well-known notions. Now consider a much less well-known operator $\del_2\!: C_k \mapsto C_{k - 2}$. This is a ``corner operator'' or a generalized boundary operator: its purpose is to take a $k$-chain $\omega^{(k)}$ and return a linear combination of all $(k - 2)$-chains that are subsets of $\omega^{(k)}$. Acting on $C_2(\Mbb, \Z_2)$, $\del_2$ is fully defined by specifying its action on all two-cells (i.e.~plaquettes):
\bel{
  \del_2 f \equiv \sum_{v \subset f} v.
}
On $C_2(\Mbb, \Z)$, the definition is
\bel{
  \del_2 f  \equiv \sum_{v \subset f} (-1)^{\sigma_v(f)} v
}
for every face $f$. See Fig.~\ref{fig del2} for an illustration of its action on all $k$-chains.

It is reasonable to define
\bel{
  \del_2 \omega^{(0)} = \del_2 \omega^{(1)} = 0
}
for every zero- or one-chain.  Then it is trivially true that $\del_2^2 = 0$ when acting on $C_2(\Mbb, \Z)$ or $C_3(\Mbb, \Z)$. If $\del_2^2 = 0$ cannot be true for any choice on orientation on a certain lattice $\Mbb$ of dimension above three, then this lattice would be unorientable in a new, generalized sense. Such lattices can only host $\Ff_2$ theories with gauge group $\Z_2$.

In addition to $\del_2$, it is also natural to define the corresponding operators $\del_{-2}$ and $\delta_2$. The former operator takes e.g.~a vertex and returns a linear combination of faces that contain that vertex. The latter operator has a simple interpretation as the symmetric double derivative $\pder{}{x^i} \pder{}{x^j}$ on cubic lattices, as will be shown in the next Section.

The literature on possible chain complexes built using $\del_2$ appears to be nonexistent; even the definition of such operators is a rare sight (see \cite{Kubica:2019}). The formal homology theory based on the $\del_2$ operator will not be developed here. Instead, the focus will be on some peculiar properties of theories whose Hamiltonians involve the generalized (co)boundary operators $\del_2$, $\del_{-2}$ and $\delta_2$.

\subsection{Pure $\Ff_2$ gauge theories} \label{subsec pure F2}

Consider a theory of Abelian fields on plaquettes of an arbitrary 3d lattice $\Mbb$, with position (magnetic) operators $Z_f$ and electric operators $X_f$.\footnote{This will be the same kind of gauge theory encountered in \eqref{dual global scheme}, except now it will be formulated on $\Mbb$ and not on $\Mbb^\vee$, in order to ease the notation.} An ordinary gauge constraint for such a theory would be defined on each link and would take the form
\bel{
  G_\ell \equiv \prod_{f \in \del_{-1} \ell} X_f = \1.
}
This can be called a type-$\Ff_1$ constraint, because the Gauss operator lives on $(k - 1)$-cells while gauge degrees of freedom live on $k$-cells (in this case, $k = 2$). This will not be the constraint imposed here. The gauge constraints of interest in this paper will be defined on each \emph{vertex}, taking the form
\bel{\label{def F2 G}
  G_v \equiv \prod_{f \in \del_{-2} v} X_f = \1.
}
This is a type-$\Ff_2$ constraint, because degrees of freedom live on $k$-cells while Gauss operators live on $(k - 2)$-cells (in this case, again, $k = 2$). The general gauge transformation $G[n]$ is
\bel{
  G[n] \equiv \prod_v G_v^{n_v} = \prod_f \prod_{v \in \del_2 f} X_f^{n_v} \equiv \prod_f X_f^{(\delta_2 n)_f}.
}
In terms of more familiar vector potentials, $Z_f \equiv \e^{\i A_f}$, conjugating $Z_f$ by $G[n]$ results in replacing
\bel{
  A_f \mapsto A_f + \frac{2\pi}{K} (\delta_2 n)_f
}
if the gauge group is $\Z_K$. In the limit of $K \gg 1$, a U(1) gauge transformation parameter
\bel{
  \lambda_v \equiv \frac{2\pi n_v}K
}
can be defined, leading to the transformation
\bel{\label{F2 gauge transf}
  A_f \mapsto A_f + (\delta_2 \lambda)_f.
}

The gauge-invariant algebra is generated by electric operators $X_f$ on faces and by magnetic operators $W_b \equiv \prod_{f \in b} Z_f$ on \emph{closed belts}. Closed belts were rather heuristically defined in Sec.~\ref{sec Xcube}, and now they can be formally defined as two-chains $b$ that satisfy\footnote{In ordinary two-form gauge theories, gauge-invariant magnetic operators (Wilson surfaces) live on two-cycles, i.e.~on two-chains $\omega^{(2)}$ that are closed, $\del \omega^{(2)} = 0$. Further, all \emph{local} magnetic operators necessarily live on two-cycles that are boundaries of three-chains (i.e.~cubes).  In $\Ff_2$ theories, however, local magnetic operators live on belts that are $\del_2$-closed but not exact.}
\bel{
  \del_2 b = 0.
}
Gauge-invariant states are closed belts of electric flux, and the theory generically has both local excitations (glueballs) and topological excitations due to flux belts winding along noncontractible one-cycles.

\begin{figure}
\begin{center}
\begin{tikzpicture}[scale = 2]

\gridThreeD{0}{0}{dotted}22;

\fill[olive!70] (0.4, 2.4) -- (0.4, 1.4) -- (0.8, 1.8) -- (0.8, 2.4) -- (0.4, 2.4);

\fill[olive!40] (0.4, 1.4) -- (0.8, 1.8) -- (1.4, 1.8) -- (1.4, 1.4) -- (0.4, 1.4);

\draw[step = 1, dotted, xshift=0.8cm, yshift=0.8cm] (0.0, 0.0) grid (2, 2);

\fill[olive] (0.4, 2.4) -- (1.4, 2.4) -- (1.4, 1.4) -- (1.8, 1.8) -- (1.8, 2.8) -- (0.8, 2.8) -- (0.4, 2.4);

\fill[blue!80] (1.4, 0.4) -- (2.4, 0.4) -- (2.8, 0.8) -- (2.8, 1.8) -- (2.4, 1.4) -- (1.4, 1.4) -- (1.4, 0.4);

\fill[blue!60] (1.8, 1.8) -- (2.8, 1.8) -- (2.4, 1.4) -- (1.8, 1.4) -- (1.8, 1.8);
\fill[blue!30] (1.8, 1.8) -- (1.8, 1.4) -- (1.4, 1.4) -- (1.8, 1.8);

\gridThreeD{0}{1}{dotted}22;
\gridThreeD{0}{2}{dotted}22;

\fill[red!60] (1.4, 1.4) -- (1.4, 0.4) -- (1, 0.4) -- (1, 1) -- (1.4, 1.4);
\fill[red!30] (1, 0) -- (1.4, 0.4) -- (1, 0.4) -- (1, 0);

\draw[step = 1, dotted, xshift=0.4cm, yshift=0.4cm] (0.0, 0.0) grid (2, 2);

\fill[red!80] (0, 0) -- (0, 1) -- (0.4, 1.4) -- (1.4, 1.4) -- (1, 1) -- (1, 0) -- (0, 0);

\draw[step = 1, dotted] (0, 0) grid (2, 2);

\draw[white] (0.6, 0.6) node {\contour{red!50!black}{$W_{c_1}^x$}};
\draw[white] (2.4, 1) node {\contour{blue!50!black}{$W_{c_3}^z$}};
\draw[white] (1.6, 2.5) node {\contour{olive!50!black}{$W_{c_2}^y$}};

\draw (4, 0.6) -- (4, 1.6) -- (5, 1.6) -- (5, 0.6) -- (4, 0.6);
\draw (4, 1.6) -- (4.4, 2) -- (5.4, 2) -- (5, 1.6);
\draw (5.4, 2) -- (5.4, 1) -- (5, 0.6);

\draw[blue] (4, 0.6) node[above right] {$+$};
\draw[blue] (5, 1.6) node[below left] {$+$};
\draw[blue] (4.05, 1.55) node[above right] {$+$};
\draw[blue] (5.35, 2.05) node[below left] {$+$};
\draw[blue] (5.475, 1.95) node[below left] {$+$};
\draw[blue] (4.95, 0.675) node[above right] {$+$};

\draw[red] (5, 0.6) node[above left] {$-$};
\draw[red] (4, 1.6) node[below right] {$-$};
\draw[red] (5.0, 1.6) node[right] {$-$};
\draw[red] (4.95, 1.55) node[above] {$-$};
\draw[red] (4.45, 2.05) node[below] {$-$};
\draw[red] (5.45, 1.05) node[left] {$-$};
\end{tikzpicture}
\end{center}
\caption{\small \emph{Left}, three different field strength operators $W^i_c$ from \eqref{def W mu} on different cubes $c_1$, $c_2$, and $c_3$ of a cubic lattice. Each of these operators is the product of four position/magnetic operators (``Wilson surfaces'') $Z_f$ that form a colored belt around the appropriate cube $c$ and that are parallel to the given direction $i$. \emph{Right}, the orientation choice used in this paper, ensuring that $\delta_2 \varphi = \del_{xy} \varphi$.
}
\label{fig field strengths}
\end{figure}
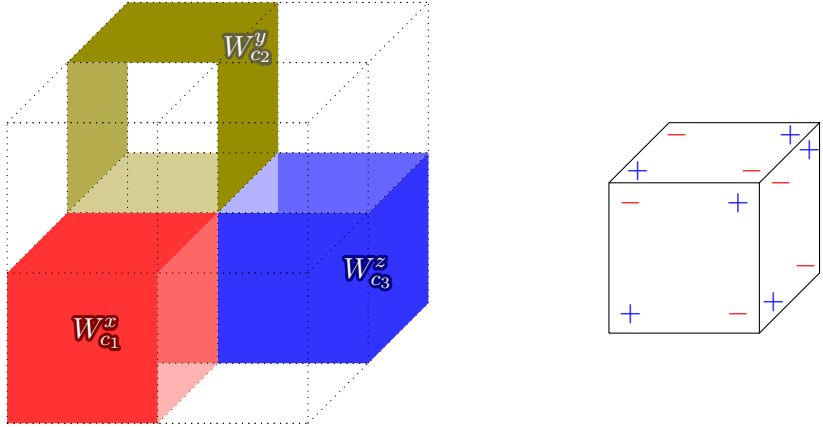

If $\Mbb$ is cubic, then every cube contains three belts, each composed of four plaquettes that are all parallel to a specific direction $i$, with magnetic operators
\bel{\label{def W mu}
  W_c^i \equiv \prod_{\substack{f \in \del c \\ f \parallel i}} Z_f.
}
See Fig.~\ref{fig field strengths}. These operators --- the field strengths of $\Ff_2$ rank-two gauge fields --- are all required to equal $\1$ in the dual X-cube model \eqref{dual global scheme}, as per the flatness condition \eqref{flatness}. In this case all electric flux belt states acquire a limited topological character: belts are locally movable in the direction perpendicular to any of their constituent  plaquettes without changing the state. Counting these states can be shown to match the ground state degeneracy in the topological phase of the X-cube model.

On more general lattices, it is again possible to impose the flatness condition \bel{
  W_b = \1
}
for any belt $b$ that winds along a contractible one-cycle. A nontrivial example will be worked out in some detail below, in the case of a body-centered-cubic lattice.

Finally, it is important to remark that $\Ff_2$ rank-two theories have an electric higher form symmetry, just like all gauge theories with a Gauss law $G = \1$ should. Unsurprisingly, this symmetry takes an unusual form: its generators are associated to one-chains that are $\del_2$-exact, or in other words to one-chains that are edges of three-chains. For instance, on a cubic lattice, this two-form symmetry is generated by operators
\bel{\label{def Ue}
  U_{e^\vee} \equiv \prod_{f \subset e^\vee} X_f
}
where $e^\vee = \del_2 c^\vee$ are edges of a dual cube, with each dual link in $e^\vee$ being dual to a plaquette in the original lattice. These generators \emph{are} topological: because of the constraint $G_c = \1$, all regions of the dual lattice that are equivalent (in $\del_2$ homology) give the same operator $U_{e^\vee}$. Thus the independent generators of this two-form symmetry are labeled by nontrivial elements of the $\del_2$ homology group $H_1(\Mbb, \Z_2)$. This should be compared to the situation in ordinary gauge theories, where the electric one-form symmetry is generated by operators living on boundaries of dual cubes, with independent generators associated to nontrivial elements of the usual $\del$ homology group $H_2(\Mbb, \Z_2)$.

\subsection{Coulomb regimes of $\Ff_2$ gauge theories}

When the gauge group is U(1), or $\Z_K$ with large but finite $K$, it makes sense to discuss the Coulomb regime of the theory. If the theory has the  Kogut-Susskind Hamiltonian
\bel{\label{def Hg}
  H_g = \frac{g^2}{2(2\pi/K)^2} \sum_f \left(2 - X_f - X_f^{-1}\right) + \frac1{2g^2} \sum_c \sum_{b \subset c} \left(2 - W_b - W_b^{-1}\right),
}
the Coulomb regime is achieved by taking $g^2$ small but still much greater than $1/K^2$.\footnote{If $g^2$ were much smaller than $1/K^2$, the theory would necessarily be in a topological phase. See \cite{Radicevic:2015sza} for a detailed discussion of this order of limits in ordinary gauge theories.} In this limit, large fluctuations of nearby gauge fields are energetically suppressed, and it makes sense to expand $Z_f \approx \1 + \i A_f$ in fluctuations $A_f$ that are $O(1/gK)$ or smaller.

It is important to keep in mind that the Coulomb regime need \emph{not} be a genuine phase --- recall that in $(2 + 1)$D, for an ordinary U(1) gauge theory, the gauge fields are confined at arbitrarily small $g^2$, with mass gap $\sim\e^{-1/g^2}$ \cite{Polyakov:1975rs}. In fact, the Coulomb regime of the $\Ff_2$ theory on the cubic lattice has already been studied in Ref.~\cite{Xu:2008} (albeit with a different UV completion); their results indeed indicate that the theory is gapped. Nevertheless, it is very useful to study the Coulomb regime because it is still applicable at all but the largest distance scales, and in its realm of applicability it gives a Gaussian theory and thereby provides an explicit bridge to familiar continuum physics.

\subsubsection{Cubic lattice}

In the Coulomb regime, the electric operators can be written as $X_f = \exp\left\{-\frac{2\pi}K \pder{}{A_f}\right\}$, leading to the Hamiltonian on the cubic lattice
\bel{\label{def H Coulomb}
  H_g \approx - \frac{g^2}2 \sum_f \pdder{}{A_f} + \frac1{2g^2} \sum_{c, \, i} \left(B_c^i \right)^2, \quad B_c^i \equiv \sum_{\substack{f \in \del c \\ f \parallel i}} A_f.
}
The vector potentials take values in increments of $2\pi/K$, and all the wavefunctions are assumed to be smooth, meaning that $\Psi[A] - \Psi[A + \frac{2\pi}K \delta^{(f)}] = O(1/K)$, with $\delta^{(f)}$ a delta-function on a face $f$.

Exploiting the cubic structure of the lattice, $A_f$ can be written as a three-component field $A_v^i$ on sites, with $i$ denoting the direction perpendicular to a given $f$. Going to the continuum notation, the fields $A_v^i$ can be written as the triplet
\bel{
  \big(A_{xy}(\b r), A_{yz}(\b r), A_{zx}(\b r)\big).
}
It is important to note that $A_{ij}(\b r)$ has no intrinsic symmetry or antisymmetry of the indices; only the three quantities above are defined on the lattice, and there is freedom to define e.g.~$A_{yx}(\b r)$ as either $A_{xy}(\b r)$ or $-A_{xy}(\b r)$ --- or as something else entirely. The gauge transformation \eqref{F2 gauge transf} is now
\bel{\label{F2 gauge transf cont}
  A_{ij} \mapsto A_{ij} + \del_{ij} \lambda.
}
This transformation makes it seem natural that $A_{ij}$ can be understood to be a symmetric tensor; the analysis below will reveal this to be a hasty conclusion, however.

The continuum magnetic fields related to local operators $W_c^i$ in \eqref{def W mu} and to the $B_c^i$ in \eqref{def H Coulomb} are
\bel{\label{def B}
  B_x = \del_y A_{zx} - \del_z A_{xy}, \quad B_y = \del_z A_{xy} - \del_x A_{yz}, \quad B_z = \del_x A_{yz} - \del_y A_{zx},
}
and the Coulomb regime Hamiltonian is
\bel{\label{def H Coulomb cont}
  H_g = \sum_{\b r} \left[- \frac{g^2}2 \left( \pdder{}{A_{xy}} + \pdder{}{A_{yz}} + \pdder{}{A_{zx}}\right) + \frac1{2g^2} \left(B_x^2 + B_y^2 + B_z^2 \right) \right].
}
Note that each magnetic field is gauge-invariant because of the symmetry of indices in the second order derivative; for instance, the gauge transformation \eqref{F2 gauge transf cont} acts on $B_x$ as
\bel{
  B_x \mapsto \del_y A_{zx} + \del_{yzx}\lambda - \del_z A_{xy} - \del_{zxy}\lambda = B_x.
}
It is again natural to choose $A_{ij}$ to be a symmetric tensor with no diagonal elements. This way one obtains the ``hollow rank-two tensor gauge theory''  whose Higgsing gives the $\Z_2$ X-cube model \cite{Ma:2018nhd}. A similar gauge theory was also argued to describe the low energy limit of a resonating plaquette model in 3d \cite{Xu:2008}.

The action corresponding to the Hamiltonian \eqref{def H Coulomb cont}, subject to the constraint \eqref{F2 gauge transf cont}, can now be found using completely standard techniques. The Lagrangian is
\bel{\label{def Lg}
  \L = \frac1{2g^2} \left[F_{0x}^2 + F_{0y}^2 + F_{0z}^2 - \left(B_x^2 + B_y^2 + B_z^2 \right)\right] \equiv \frac1{2g^2} F_{\mu\nu}F^{\mu\nu},
}
summing over the unusual index $\mu\nu \in \{xy, yz, zx, 0x, 0y, 0z \}$, and with the field strengths
\gathl{\label{def F}
  F_{0x} \equiv \del_0 A_{yz} - \del_{yz} A_0, \quad F_{0y} \equiv \del_0 A_{zx} - \del_{zx} A_0, \quad F_{0z} \equiv \del_0 A_{xy} - \del_{xy} A_0,\\
  F_{xy} \equiv B_z, \quad F_{yz} \equiv B_x, \quad F_{zx} \equiv B_y.
}
(The fields $B_i$ were defined in \eqref{def B}.) The gauge transformation of the time component of the vector potential is
\bel{
  A_0 \mapsto A_0 + \del_0 \lambda.
}
In every respect this component behaves like an ordinary, one-form gauge field; this is inherited from the fact that the gauge constraints are defined on vertices, not links. In particular, it is possible to gauge-fix away this component by going to temporal gauge, $A_0 = 0$, whereby the field strengths reduce to electric fields, $F_{0x} = \del_0 A_{yz}$ and so on.

The equation of motion for $A_0$ is, of course, just the Gauss law $\del_{xy} F_{0z} + \del_{yz} F_{0x} + \del_{zx} F_{0y} = 0$. More interesting are the equations for the spatial components of vector potentials $A_{ij}$,
\bel{\label{EoM}
  \del_0 F_{0z} = \del_z (B_y - B_x), \quad \del_0 F_{0y} = \del_y (B_x - B_z), \quad \del_0 F_{0x} = \del_x (B_z - B_y).
}
Consider the following simple yet surprising fact concerning the (anti)symmetry of indices. Recall that $\del_0 F_{0z} = \del_{00} A_{xy} - \del_{0xy} A_0$ should be symmetric in $x$ and $y$, if $A_{ij}$ is chosen to be symmetric due to its gauge transformation law \eqref{F2 gauge transf cont}. On the other hand, the r.h.s.~of the above equation is evidently antisymmetric under exchanging $x$ and $y$. This antisymmetry is inherited from demanding gauge invariance of the field strengths \eqref{def B}. The conclusion is that there is, in fact, \emph{no} natural way to think of the (anti)symmetry properties of $A_{ij}$. At the end of the day, only $A_{xy}$, $A_{yz}$, and $A_{zx}$ are ever defined in the UV, and there is no reason why well-established intuition from ordinary differential geometry should apply to $\Ff_2$ theories.

Another observation is that eq.~\eqref{def Lg} appears to be a Gaussian theory with more than two derivatives in the action, but all higher derivative terms disappear in the temporal gauge $A_0 = 0$. In this gauge, for example, the first equation of motion in \eqref{EoM} is
\bel{
  (-\del_{00} + 2\del_{zz}) A_{xy} = \del_{zy} A_{zx} + \del_{zx} A_{yz}.
}
The other equations are obtained by cyclic permutations $x \mapsto y \mapsto z \mapsto x$.

The spectrum is easy to obtain by expressing the above equation of motion in momentum space. As in ordinary gauge theories, one of the three gauge field components is not dynamical, so there are only two physical polarizations of the photon. They have the unusual dispersion \cite{Xu:2008}
\bel{\label{photon disp}
  \omega^2_\pm =  p_x^2 + p_y^2 + p_z^2 \pm \sqrt{p_x^4 + p_y^4 + p_z^4 - p_x^2 p_y^2 - p_y^2 p_z^2 - p_z^2 p_x^2}.
}
A striking property of these photons is that they have limited mobility (requirement F2) in the following sense: if $p_y = p_z = 0$, say, one component of the polarization becomes static, with dispersion $\omega_- = 0$ for any $p_x$, while the ``+'' polarization propagates alone along the $x$ axis with $\omega_+ = \sqrt 2 p_x$. The ground state thus has an exponential degeneracy characteristic of fracton theories (requirement F1). It would be interesting to find out whether this degeneracy/restricted mobility combination survives at the nonperturbative level.

As argued even before passing to the continuum, this theory possesses infinitely many electric higher-form symmetries. However, recall that in ordinary gauge theories, the Coulomb regime displayed an emergent magnetic higher-form symmetry. In the continuum, this extra conservation came from the Bianchi identity, and it essentially amounted to saying that the magnetic flux through a surface must be constant if all the fields are smooth. In the case of theories studied here, a similar argument can be run to show that there must exist a \emph{magnetic two-form symmetry} for each plane $p$. Essentially, if the fields vary smoothly, then $W_b^{i}$ over a contractible belt must be trivial because the belt can be shrunk to zero size, and this can be done in each plane separately. Thus pure $\Ff_2$ U(1) rank-two theories have an extensive number of magnetic two-form symmetries, one for each plane. Monopole excitations in these theories exist within each plane separately, but all are infinitely heavy in the continuum limit; they do not contribute to the exponential degeneracy of the ground state.

\subsubsection{BCC lattice} \label{subsec BCC}

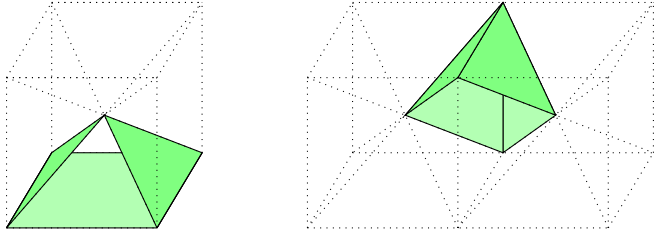
\begin{figure}
\begin{center}
\begin{tikzpicture}[scale = 2]

\draw[dotted, step = 1, xshift=0.3cm, yshift=0.5cm] (0.0, 0.0) grid (1,1);

\filldraw[fill = green!30] (0, 0) -- (1, 0) -- (1.3, 0.5) -- (0.3, 0.5) -- (0, 0);
\filldraw[fill = green!50] (0, 0) -- (0.65, 0.75) -- (0.3, 0.5) -- (0, 0);
\filldraw[fill = green!50] (1, 0) -- (0.65, 0.75) -- (1.3, 0.5) -- (1, 0);

\draw(0, 0) -- (0.3, 0.5);
\draw (1, 0) -- (1.3, 0.5);

\draw[dotted, step = 1] (0, 0) grid (1,1);

\gridThreeD{0}{1}{dotted}11;

\draw[dotted] (0, 0) -- (1.3, 1.5);
\draw[dotted] (0, 1) -- (1.3, 0.5);
\draw[dotted] (1, 0) -- (0.3, 1.5);
\draw[dotted] (1, 1) -- (0.3, 0.5);

%%%%%%%%

\draw[step = 1, xshift=2.3cm, yshift=0.5cm, dotted] (0.0, 0.0) grid (2,1);

\filldraw[fill = green!30] (3.3, 1.5) -- (3.3, 0.5) -- (3.65, 0.75) -- (3.3, 1.5);
\filldraw[fill = green!30] (3.3, 1.5) -- (3.3, 0.5) -- (2.65, 0.75) -- (3.3, 1.5);

\filldraw[fill = green!50] (3.65, 0.75) -- (3.3, 1.5) -- (3, 1) -- (3.65, 0.75);
\filldraw[fill = green!50] (3.3, 1.5) -- (2.65, 0.75) -- (3, 1) -- (3.3, 1.5);

%\fill[green!30] (0, 0) -- (1, 0) -- (1.3, 0.5) -- (0.3, 0.5) -- (0, 0);
%\fill[green!50] (0, 0) -- (0.65, 0.75) -- (0.3, 0.5) -- (0, 0);
%\fill[green!50] (1, 0) -- (0.65, 0.75) -- (1.3, 0.5) -- (1, 0);

\draw[dotted] (2, 0) -- (2.3, 0.5);
\draw[dotted] (3, 0) -- (3.3, 0.5);
\draw[dotted] (4, 0) -- (4.3, 0.5);

\draw[dotted, step = 1] (2, 0) grid (4,1);

\draw[dotted] (2, 1) -- (2.3, 1.5);
\draw[dotted] (3, 1) -- (3.3, 1.5);
\draw[dotted] (4, 1) -- (4.3, 1.5);

\draw[dotted] (2, 0) -- (3.3, 1.5);
\draw[dotted] (2, 1) -- (3.3, 0.5);
\draw[dotted] (3, 0) -- (2.3, 1.5);
\draw[dotted] (3, 1) -- (2.3, 0.5);

\draw[dotted] (3, 0) -- (4.3, 1.5);
\draw[dotted] (3, 1) -- (4.3, 0.5);
\draw[dotted] (4, 0) -- (3.3, 1.5);
\draw[dotted] (4, 1) -- (3.3, 0.5);

\end{tikzpicture}
\end{center}
\caption{\small \emph{Left}: the unit cell of a BCC lattice. The fields $A^{ab}_{i}(\b r)$ live on triangular faces, while $A_{ij}(\b r)$ live on square faces.  Elementary field strengths $B^{ab}_{ij}(\b r)$ live on tents, i.e.~on two-chains built out of one square plaquette and two triangular ones. One tent is shown above shaded in green. \emph{Right}: plaquettes containing one of the additional ``two-cell'' magnetic field operators, which are for simplicity assumed to be tuned away.
}
\label{fig BCC}
\end{figure}

The BCC (body-centered-cubic) lattice is obtained by starting with a cubic lattice and adding a site to the center of each cube, with links connecting this site to all six vertices of the cube. Thus the BCC lattice has all the square faces of the cubic lattice \emph{and} an additional 12 triangular faces per cube, connecting the cube links to its center. These faces partition each cube into six pyramidal cells. See Fig.~\ref{fig BCC} for an illustration.

\renewcommand{\xytransf}[2]
{ % first 4 params = transf matrix that changes the coordinate system
  % 5th param = origin of the transformation
    \pgftransformcm{1}{0}{0.3}{0.5}{\pgfpoint{#1cm}{#2cm}}
}

\renewcommand{\yztransf}[2]
{
   \pgftransformcm{0.3}{0.5}{0}{1}{\pgfpoint{#1cm}{#2cm}}
}

Now consider the pure, U(1), $\Ff_2$ rank-two gauge theory on this lattice. The gauge constraint \eqref{def F2 G} can be defined on this lattice too. However, to pass to continuum notation in the Coulomb regime, it is necessary to supplement the three fields $A_{ij}(\b r)$ with another 12 fields. These can be denoted by $A_i^{ab}(\b r)$ via the following convention. For a given vertex of the cubic lattice that corresponds to $\b r$, first pick a link emanating from this vertex in direction $i$ for $i \in \{x, y, z\}$. Then, pick one of the four triangular BCC faces that contain this link. These can be labeled by $a, b \in \{+, -\}$. The center of the cube corresponding to each of these four choices can be denoted $\b r + \frac12 \b e_i + \frac12 \Delta_i^{ab}$, where
\bel{\label{def Delta x ab}
  \Delta_x^{ab} \equiv a \b e_y  +  b \b e_z,
}
with $\b e_i$ being unit vectors on the original cubic lattice, and with $\Delta_{y/z}^{ab}$ obtained by cyclically permuting the indices above.

\begin{figure}
\begin{center}
\begin{tikzpicture}[scale = 2]

\draw[dotted, step = 1, xshift=0.3cm, yshift=0.5cm] (0.0, 0.0) grid (1,1);

%\filldraw[fill = green!30] (0, 0) -- (1, 0) -- (1.3, 0.5) -- (0.3, 0.5) -- (0, 0);
\filldraw[fill = green!30] (0, 0) -- (0.65, 0.75) -- (0.3, 0.5) -- (0, 0);
\filldraw[fill = green!30] (1, 0) -- (0.65, 0.75) -- (1.3, 0.5) -- (1, 0);
\filldraw[fill = green!30] (0, 1) -- (0.65, 0.75) -- (0.3, 1.5) -- (0, 1);
\filldraw[fill = green!30] (1, 1) -- (0.65, 0.75) -- (1.3, 1.5) -- (1, 1);

\draw(0, 0) -- (0.3, 0.5);
\draw (1, 0) -- (1.3, 0.5);

\draw[dotted, step = 1] (0, 0) grid (1,1);

\gridThreeD{0}{1}{dotted}11;

\draw[dotted] (0, 0) -- (1.3, 1.5);
\draw[dotted] (0, 1) -- (1.3, 0.5);
\draw[dotted] (1, 0) -- (0.3, 1.5);
\draw[dotted] (1, 1) -- (0.3, 0.5);

\draw[blue] (0, 0) node[left] {$\boldsymbol +$};
\draw[blue] (1, 0) node[right] {$\boldsymbol +$};
\draw[blue] (0, 1) node[left] {$\boldsymbol +$};
\draw[blue] (1, 1) node[right] {$\boldsymbol +$};

\draw[red] (0.3, 0.5) node[left] {$\boldsymbol -$};
\draw[red] (1.3, 0.5) node[right] {$\boldsymbol -$};
\draw[red] (0.3, 1.5) node[left] {$\boldsymbol -$};
\draw[red] (1.3, 1.5) node[right] {$\boldsymbol -$};

\draw[purple] (0.65, 0.75) node {\contour{white}{$\boldsymbol \pm$}};

%%%

\begin{scope}[shift = {(2.5,0)}]
\draw[dotted, step = 1, xshift=0.3cm, yshift=0.5cm] (0.0, 0.0) grid (1,1);

\filldraw[fill = green!30] (0.3, 1.5) -- (0.65, 0.75) -- (0.3, 0.5) -- (0.3, 1.5);
\filldraw[fill = green!30] (1.3, 1.5) -- (0.65, 0.75) -- (1.3, 0.5) -- (1.3, 1.5);
\filldraw[fill = green!30] (0, 0) -- (0.65, 0.75) -- (0, 1) -- (0, 0);
\filldraw[fill = green!30] (1, 0) -- (0.65, 0.75) -- (1, 1) -- (1, 0);

%\draw[dotted] (0, 0) -- (0.3, 0.5);
\draw[dotted] (1, 0) -- (1.3, 0.5);
%\draw[dotted] (0.3, 0.5) -- (0.3, 1.5);

\draw[dotted, step = 1] (0, 0) grid (1,1);

\gridThreeD{0}{1}{dotted}11;

\draw[dashed] (0.3, 0.5) -- (0.3, 1.5);
\draw[dashed] (0.3, 0.5) -- (0.65, 0.75);
%\draw[dotted] (1, 0) -- (0.3, 1.5);
%\draw[dotted] (1, 1) -- (0.3, 0.5);

\draw[blue] (0, 0) node[left] {$\boldsymbol +$};
\draw[blue] (1, 0) node[right] {$\boldsymbol +$};
\draw[red] (0, 1) node[left] {$\boldsymbol -$};
\draw[red] (1, 1) node[right] {$\boldsymbol -$};

\draw[blue] (0.3, 0.5) node[left] {$\boldsymbol +$};
\draw[blue] (1.3, 0.5) node[right] {$\boldsymbol +$};
\draw[red] (0.3, 1.5) node[left] {$\boldsymbol -$};
\draw[red] (1.3, 1.5) node[right] {$\boldsymbol -$};

\draw[purple] (0.65, 0.75) node {\contour{white}{$\boldsymbol \pm$}};
\end{scope}

\begin{scope}[shift = {(5, 0)}]
\draw[dotted, step = 1, xshift=0.3cm, yshift=0.5cm] (0.0, 0.0) grid (1,1);

\filldraw[fill = green!30] (0.3, 0.5) -- (0.65, 0.75) -- (1.3, 0.5) -- (0.3, 0.5);
\filldraw[fill = green!30] (0.3, 1.5) -- (0.65, 0.75) -- (1.3, 1.5) -- (0.3, 1.5);
\filldraw[fill = green!30] (0, 0) -- (0.65, 0.75) -- (1, 0) -- (0, 0);
\filldraw[fill = green!30] (0, 1) -- (0.65, 0.75) -- (1, 1) -- (0, 1);

\draw[dotted] (0, 0) -- (0.3, 0.5);
\draw[dotted] (1, 0) -- (1.3, 0.5);

\draw[dotted, step = 1] (0, 0) grid (1,1);

\gridThreeD{0}{1}{dotted}11;

%\draw[dotted] (0, 0) -- (1.3, 1.5);
\draw[dotted] (0, 1) -- (1.3, 0.5);
%\draw[dotted] (1, 0) -- (0.3, 1.5);
\draw[dotted] (1, 1) -- (0.3, 0.5);

\draw[blue] (0, 0) node[left] {$\boldsymbol +$};
\draw[red] (1, 0) node[right] {$\boldsymbol -$};
\draw[blue] (0, 1) node[left] {$\boldsymbol +$};
\draw[red] (1, 1) node[right] {$\boldsymbol -$};

\draw[blue] (0.3, 0.5) node[left] {$\boldsymbol +$};
\draw[red] (1.3, 0.5) node[right] {$\boldsymbol -$};
\draw[blue] (0.3, 1.5) node[left] {$\boldsymbol +$};
\draw[red] (1.3, 1.5) node[right] {$\boldsymbol -$};

\draw[purple] (0.65, 0.75) node {\contour{white}{$\boldsymbol \pm$}};
\end{scope}

\end{tikzpicture}
\end{center}
\caption{\small A choice of orientations of the triangular plaquettes in the BCC lattice. (The orientations of square plaquettes are as shown in Fig.~\ref{fig field strengths}.) There are twelve triangular plaquettes per cube; here their orientations are shown in groups of four, to improve legibility. The signs in the corners indicate the orientations of vertices that form the green-shaded plaquettes. It is with these signs that the corresponding vertices appear in the zero-chain $\del_2 f$ for each green $f$, and these are also the signs with which $\lambda_v$'s appear in the differential $(\delta_2 \lambda)_f$. \emph{All} triangular plaquettes have the same orientation in the center of the cube, which can be either $+$ or $-$. The orientation of this central vertex can be chosen to be constant across the lattice, or it can vary from cube to cube.
}
\label{fig BCC orient}
\end{figure}
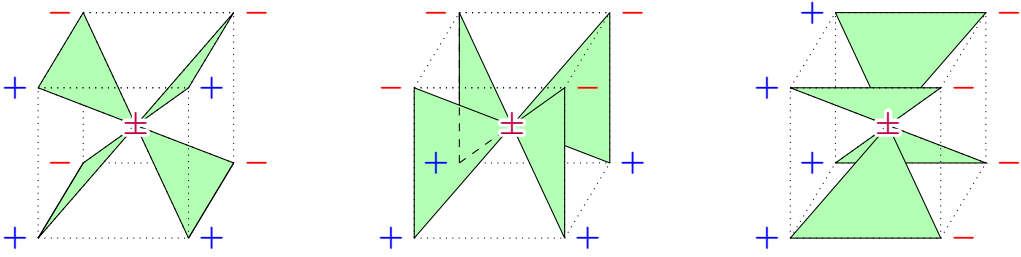

There are now two gauge parameters for each $\b r$, one corresponding to a scalar field on original cube vertices, and one corresponding to a scalar field on cube centers. The former can still be denoted $\lambda(\b r)$, while the latter can be denoted by $\lambda\_c(\b r) \equiv \lambda\left(\b r + \frac12 \b e_x + \frac12 \b e_y + \frac12 \b e_z\right)$. Taking into account the orientations shown on Fig.~\ref{fig BCC orient}, the gauge transformations are
\gathl{\label{def BCC G}
  A_{ij}(\b r) \mapsto A_{ij}(\b r)  + \del_{ij} \lambda(\b r),\\
  A_i^{ab}(\b r) \mapsto A_i^{ab}(\b r) - \del_i \lambda(\b r) + \lambda\left(\b r + \frac12 \b e_i + \frac12 \Delta_i^{ab}\right).
}
Here and below, every term of the form $\phi(\b r + \b n)$ can be equivalently written as $\phi(\b r) + n^i \del_i\phi(\b r)$ for an \emph{integer} vector $\b n$. A half-integer shift of $\lambda(\b r)$ gives $\lambda\_c$. In particular, it will be useful to let
\bel{\label{def tilde Delta}
  \lambda\left(\b r + \frac12 \b e_i + \frac12 \Delta_i^{ab}\right) \equiv \lambda\_c\big(\b r + \~\Delta_i^{ab}\big).
}
%This definition of $\~\Delta^{ab}_i$ will be used below.

The magnetic field also has multiple components. There is one for each local tent (see Fig.~\ref{fig BCC}), labeled by $B_{ij}^{ab}(\b r)$ with the following meaning. Starting from the vertex labeled $\b r$, consider the three plaquettes on which fields $A_{xy}(\b r)$, $A_{yz}(\b r)$, and $A_{zx}(\b r)$ live. Pick one of these plaquettes, and label it by $ij$. Next, pick one side of this plaquette; label this choice by $a \in \{+, -\}$. The coordinates of the cube centers on either side of the plaquette can be denoted $\b r + \frac12\b e_i + \frac12 \b e_j + \frac12  a \b e_k$, with $k \neq i, j$.

There are two belts that contain any chosen plaquette $ij$ and the cube center on the chosen side $a$ of the plaquette.  Each of these belts contains two triangular faces that share the cube center while not sharing any links. These belts can be labeled by $b \in \{+, -\}$. These are the \emph{tents} shown on Fig.~\ref{fig BCC}. The magnetic field on the four possible tents constructed above the square plaquette $ij$ at site $\b r$ are, for example,
\algns{
  B_{yz}^{++} (\b r)
  &\equiv A_{yz}(\b r) - A_y^{++}(\b r) + A_y^{-+}(\b r + \b e_z),\\
  B_{yz}^{+-} (\b r)
  &\equiv A_{yz}(\b r) - A_z^{++}(\b r) + A_z^{+-}(\b r + \b e_y),\\
  B_{yz}^{-+} (\b r)
  &\equiv A_{yz}(\b r) - A_y^{+-}(\b r) + A_y^{-+}(\b r + \b e_z),\\
  B_{yz}^{--} (\b r)
  &\equiv A_{yz}(\b r) - A_z^{-+}(\b r) + A_z^{+-}(\b r + \b e_y).
}
The other magnetic fields $B_{ij}^{ab}(\b r)$ can be constructed analogously, using only the orientations indicated on Figs.~\ref{fig BCC} and \ref{fig BCC orient}.

It is straightforward to verify that these magnetic fields are invariant under gauge transformations \eqref{def BCC G}. Here is a sample calculation. The transformation of $B_{yz}^{++}(\b r)$ is
\bel{
  B_{yz}^{++}(\b r) \mapsto B_{yz}^{++}(\b r) + \del_{yz}\lambda(\b r) + \del_y \lambda(\b r) - \lambda\_c\big(\b r + \~\Delta^{++}_y \big) - \del_y \lambda(\b r + \b e_z) + \lambda\_c\big(\b r + \b e_z + \~\Delta^{-+}_y\big) = B_{yz}^{++}(\b r),
}
because $\del_y \lambda(\b r) - \del_y \lambda(\b r + \b e_z) = - \del_{yz} \lambda(\b r)$ and $\b e_z + \~\Delta_y^{-+} = \~\Delta_y^{++}$, as follows from eqs.~\eqref{def Delta x ab} and \eqref{def tilde Delta}.

%It is  also easy to check that summing field strengths on four tents whose bases form a belt around a cube gives back the magnetic field $B_i(\b r)$ on that belt; the nontrivial part of this exercise is to consistently assign signs when defining $B_{yz}^{+-}$ and so on.  The reader is invited to construct the remaining field strengths as an exercise.

This theory of gauge fields is interesting but notationally complicated, especially due to the lack of rotational covariance. However, one already apparent feature is rather unusual, as it is not present in any $\Ff_1$ gauge theories --- and it was also not present in the cubic case studied above. Consider a gauge transformation \eqref{def BCC G} associated to a \emph{constant} gauge parameters $\lambda(\b r) = \lambda\_c(\b r) \equiv \Lambda$. This choice does \emph{not} leave the gauge fields unchanged. The fields on triangular faces shift by a term that is not a total derivative, so the constant gauge transformation acts as
\bel{\label{nonconst gauge transf}
  A_i^{ab}(\b r) \mapsto A_i^{ab}(\b r) + \Lambda.
}

Thus this is an example of a theory in which the gauge transformations \emph{cannot} be simply viewed as generalizations of global symmetry transformations to locally varying symmetry parameters. The new ``global'' transformations must be defined as those in which $\delta_2 \lambda = 0$ holds on the underlying lattice. On a cubic lattice, such ``global'' transformations have $\del_{ij} \lambda = 0$, and this equation is solved by any function of a single variable. On the BCC lattice, in addition to $\del_{ij} \lambda = 0$ there is also the requirement that $\del_i \lambda(\b r) = \lambda\_c \big(\b r + \~\Delta^{ab}_i\big)$. This will have important implications for the kinds of matter systems that can be coupled to the $\Ff_2$ theory on the BCC lattice, as will be explained in the next Subsection.

A pair of adjacent cells can host operators that are products of $Z_f$'s (or sums of $A_f$'s) over four plaquettes (see Fig.~\ref{fig BCC}). Including all these operators in the Hamiltonian will allow belts of electric flux to twist and snap, reducing the GSD to that of a toric code. The ``fracton order'' on the BCC lattice is thus unstable. After tuning these operators so that they are not present in the Hamiltonian, topological $\Ff_2$ theories will enjoy an exponential GSD. The exact scaling differs from the cubic case because the electric flux belts can still twist, i.e.~the flux belt can pass through triangular plaquettes within the same cube that do not have parallel long sides. This means that there exist gauge-invariant states in which flux belts change direction, with the points at which the direction changes being immobile (no operator allowed in the Hamiltonian can create or move these ``defects''). In the sector of the theory without any such defects, all the electric flux lines remain confined to a single plane, and the GSD of the cubic lattice is recovered.

\subsection{$\Ff_2$ theories coupled to matter}

The dual X-cube model is not a pure $\Ff_2$ theory, as it contains $\Z_2$ matter fields too. The purpose of this Section is to generalize the X-cube model to arbitrary gauge groups and lattices by incorporating matter (i.e.~dynamical charges) into the above general discussion of $\Ff_2$ gauge theories. The dynamics --- and in particular, the infrared behavior --- of such theories for general gauge groups remains far from understood, though see \cite{Chamon:2005, Prem:2017qcp, Weinstein:2018xil} for some results.

It is trivial to add classical (nondynamical) background fields to the $\Ff_2$ theory with Gauss operators \eqref{def F2 G}, for a general gauge group $\Z_K$. The gauge constraint merely needs to be replaced by
\bel{
  \varrho_v G_v \equiv \varrho_v \prod_{f \in \del_{-2} v} X_f =  \1,
}
for any function $\varrho\!: C_0 \rar \Z_K$.

To make the matter dynamical, the background density $\varrho_v$ is replaced by the matter momentum operator $X_v$. (In field theory language, this operator generates the rotations of a complex field $\Phi \mapsto \e^{\i \lambda} \Phi$, or shifts of a compact scalar $\varphi \mapsto \varphi + \lambda$.) The gauge constraint is
\bel{
  X_v G_v = \1.
}
All matter momentum operators are gauge-invariant, but the position operators $Z_v$ must come in combinations defined on plaquettes,
\bel{\label{def Sf}
  S_f \equiv Z_f^{-1} \prod_{v \in \del_2 f} Z_v.
}
This is the $\Ff_2$ generalization of the familiar kinetic term in ordinary gauge theories, where individual position operators are not gauge-invariant, and instead one works with gauge-invariant combinations on links, $S_\ell = Z_\ell^{-1} \prod_{v \in \del \ell} Z_v$. If the gauge group is U(1), the ordinary kinetic term can be rewritten using $Z_\ell = \e^{\i A_\ell}$ and $Z_v = \e^{\i \varphi_v}$ to become $S_\ell = \e^{\i (\delta \varphi - A)_\ell}$ --- this is simply the exponential of  (a version of) the covariant derivative on the link $\ell$. Thus the plaquette operator $S_f$ in \eqref{def Sf} contains the $\Ff_2$ analogue of the covariant derivative, and in the U(1) case it can be written as
\bel{\label{Sf alternate}
  S_f = \e^{\i (\delta_2 \varphi - A)_f}.
}
The plaquette variable $A_f$ is the ``Wilson membrane,'' generalizing the notion of a gauge connection.

The natural Hamiltonian for the matter fields (in the limit in which large fluctuations of all fields are suppressed) is
\algns{\label{def Hm}
  H\_m
  &= \frac{e^2}2 \sum_f \left(2 - S_f - S^{-1}_f\right) + \frac{u^2}{2(2\pi/K)^2} \sum_v \left(2 - X_v - X_v^{-1}\right) \\
  &\approx \frac{e^2}2 \sum_f \left((\delta_2 \varphi)_f - A_f\right)^2  - \frac{u^2}{2} \sum_v \pdder{}{\varphi_v}.
}
Quantum field theorists would recognize this as the St\"uckelberg coupling between gauge and matter fields. A more familiar theory of a complex field coupling to the gauge field, with a kinetic term of the form $|(\delta_2 - A) \Phi|^2$, is not a natural object on the lattice. It may well be the case that such an object will prove interesting to analyze from the continuum point of view, however.

\subsubsection{Cubic lattice}

On a cubic lattice, the previously used convention for labeling gauge fields as $A_{ij}(\b r)$ can be employed here. The matter part of the Hamiltonian then becomes
\bel{
  H\_m = \sum_{\b r} \left[\frac{e^2}2 \left((\del_{xy} \varphi - A_{xy})^2 + (\del_{yz} \varphi - A_{yz})^2 + (\del_{zx} \varphi - A_{zx})^2 \right)  - \frac{u^2}2 \pdder{}{\varphi} \right],
}
with matter transforming under gauge transformations \eqref{F2 gauge transf cont} as
\bel{
  \varphi(\b r) \mapsto \varphi(\b r) + \lambda(\b r).
}

The most important property of this matter system is that the kinetic term has four-derivative terms. (This applies only to spatial derivatives; the action will still be second-derivative in time.) Without gauge fields, the matter Lagrangian would simply be
\bel{\label{def Lm}
  \L\_m = \frac{e^2}2 (\del_{ij} \varphi)(\del^{ij} \varphi) - \frac1{2u^2} (\del_0\varphi)^2,
}
with an implied summation over $ij \in \{xy, yz, zx\}$. This is an example of a theory \emph{quadratic} in fields and  invariant under the transformations
\bel{\label{linear transf}
  \varphi(\b r, t) \mapsto \varphi(\b r, t) + \Lambda + c^i r_i, \quad i \in \{x, y, z\}.
}
Theories with symmetries labeled by a vector (in this case $c^i$) were recently discussed in \cite{Pretko:2018jbi, Seiberg:2019vrp, Shenoy:2019}. Of course, these vector symmetries are just a small part of the entire family of symmetries the system \eqref{def Lm} has. For each spatial plane there is a conserved quantity, originally defined in eq.~\eqref{def Q vee}:
\bel{\label{def Q p}
  Q_p = \prod_{v \subset p} X_v = \exp\left\{-\frac{2\pi}K \sum_{v \subset p} \pder{}{\varphi_v}\right\}.
}
These are nontopological one-form symmetries, as they are defined on manifolds of spatial codimension one. The shift by a constant $\Lambda$ is generated by the product of $Q_p$ over all planes parallel to any given direction. The shift by $c^x r_x$ is enacted by $\prod_{v = 1}^{L_x} Q_{p(v)}^{c^x v}$, where $p(v)$ denotes the $yz$ plane at $x$-coordinate $v$. From the point of view of the X-cube model and its duals, the transformations \eqref{linear transf} are not special at all. The most general ``global'' symmetry of this theory that can be built out of the $Q_p$'s is
\bel{\label{varphi transf}
  \varphi(\b r, t) \mapsto \varphi(\b r, t) + \lambda_x(x) + \lambda_y(y) + \lambda_z(z),
}
with $\b r = (x, y, z)$ and for arbitrary functions of a single variable $\lambda_i$ satisfying the periodicity condition $\lambda_i(r_i + L_i) = \lambda_i(r_i)$. This is the most general transformation satisfying $\del_{ij} \lambda(\b r) = 0$, which is the requirement derived at the end of the previous Section. Similar theories were also studied in Refs.~\cite{Paramekanti:2002, Xu:2007}, although in very different contexts.

The one-form symmetries \eqref{def Q p} are gauged by introducing the now-familiar $\Ff_2$  rank-two gauge fields $A_{ij}(\b r)$.\footnote{At no point in this discussion is there a separate gauging of ``global'' symmetries associated to $\varphi \mapsto \varphi + \Lambda$ and to $\varphi \mapsto \varphi + c^i r_i$, as proposed in \cite{Wang:2019aiq}. Indeed, once arbitrary local transformations $\lambda(\b r)$ are introduced, there is no way to separate out these two types of ``global'' transformations.

It is, however, possible to gauge just the symmetry associated to a single generator $Q_p$. The gauge fields needed to do so would live \emph{only on links that belong to the plane $p$}. Introducing gauge fields needed to make \emph{every} $Q_p$ be in the singlet sector, one gets precisely the X-cube model. It is a special property of the cubic lattice that the same symmetries could have been gauged by introducing rank-two gauge fields with the $\Ff_2$ type gauge constraints.

Further, it is possible to demand that only certain combinations of different $Q_p$'s be gauged. For instance, if the goal is to gauge just $\prod_{v = 1}^{L_x} Q_{p(v)}$, this can be done by introducing ordinary gauge fields on all links of the lattice. To gauge $\prod_{v = 1}^{L_x} Q_{p(v)}^v$, the matter must be given a position-dependent charge (fields at $x$-coordinate $v$ must have charge $v$), but otherwise ordinary gauge fields on links still do the trick. This way one may rigorously construct a gauge theory (with multiple different gauge fields) that corresponds to gauging the few particular combinations of $Q_p$'s that enact the linear transformation $\varphi \mapsto \varphi + \Lambda + c^i r_i$.} But even without the gauging, matter given by the Lagrangian \eqref{def Lm} has restricted mobility due to the huge amount of symmetry. The dispersion relation of this theory is
\bel{
  \omega^2 \sim p_x^2 p_y^2 + p_y^2 p_z^2 + p_z^2 p_x^2,
}
and so no propagating excitation can have zero momenta in any two directions simultaneously. In other words, any excitation must move in at least two directions. (Compare this to the $\Ff_2$ photon dispersion \eqref{photon disp}.) Thus the pure matter theory \eqref{def Lm} satisfies conditions F2 and F3.

The ground state has a huge degeneracy (condition F1). Every field configuration that is constant along two of the three directions has zero energy. The degeneracy thus grows exponentially with the linear size of the three-manifold. Picking a particular vacuum corresponds to spontaneously breaking the subdimensional symmetries. Importantly, this is \emph{not} the same kind of degeneracy found in the X-cube model: there, the global symmetry of the matter system is gauged, the degenerate matter states are identified, and any degeneracy comes from the gauge sector.

This discussion demonstrates that the matter theory from \eqref{def Hm} on a cubic lattice, described by the Lagrangian \eqref{def Lm}, satisfies all three hallmarks of fracton physics. No gauge fields are needed to satisfy these requirements. Of course, it would be interesting to understand if nonperturbative effects may cause a gap to appear and destroy fracton behavior.

\subsubsection{BCC lattice}

On a BCC lattice, the appearance of gauge transformations without derivatives \eqref{def BCC G} in the pure $\Ff_2$ gauge theory adds an important complication to the continuum description of the $\Ff_2$ gauge theory coupled to matter. The matter theory that naturally couples to $\Ff_2$ gauge fields has a Hamiltonian built out of operators $S_f$ and $X_v$, just like its cubic lattice version \eqref{def Hm}. However, the generalized differential $(\delta_2 \varphi)_f$ that features in $S_f$ (eq.~\eqref{def Sf}) can no longer be interpreted as a double derivative when $f$ is triangular. Consider the case of a triangular plaquette $f$ whose long side is parallel to the $x$-axis.  Then $S_f$ is the exponential of the covariant derivative, as in eq.~\eqref{Sf alternate}, and its matter part is
\bel{
  (\delta_2 \varphi)_f = -\del_x \varphi(\b r) \pm \varphi\left(\b r + \frac12\b e_x + \frac12 \b e_y + \frac12 \b e_z\right) \equiv -\del_x \varphi(\b r) \pm \varphi\_c(\b r).
}
The field $\varphi\_c(\b r)$ that lives in the center of the cube enters $\delta_2\varphi$ with a sign that depends on the choice of orientation. Fig.~\ref{fig BCC orient} shows a choice of orientation in which this sign is the same for all triangular plaquettes within the same cube.

Regardless of the choice of orientation, however, $\varphi\_c$ enters $\delta_2\varphi$ \emph{without} a derivative. This means that a shift of all scalars by a constant,
\bel{
  \varphi(\b r) \mapsto \varphi(\b r) + \Lambda, \quad \varphi\_c(\b r) \mapsto \varphi\_c(\b r) + \Lambda,
}
is \emph{not} a symmetry of this theory as long as the Hamiltonian contains individual operators $S_f$. In other words, the U(1) charge of the compact scalar on a BCC lattice is not conserved for arbitrary Hamiltonians built out of $S_f$ and $X_v$.\footnote{This was, of course, also evident before passing to continuum notation. The lattice kinetic operator for the pure matter theory is $S_f = \prod_{v \in \del_2 f} Z_v$, which is obtained from eq.~\eqref{def Sf} by simply removing the gauge field operators. This operator does not commute with the total U(1) charge, $\prod_{v \in \Mbb} X_v$.} This is an analogue of the fact that pure $\Ff_2$ gauge fields on the BCC lattice are not invariant under gauge transformations with a constant gauge parameter, as shown in eq.~\eqref{nonconst gauge transf}.

The fields on the cubic sublattice do have a conserved charge, i.e.~the transformation
\bel{
  \varphi(\b r) \mapsto \varphi(\b r) + \Lambda, \quad \varphi\_c(\b r) \mapsto \varphi\_c(\b r)
}
is a symmetry for any Hamiltonian. However, it is not difficult to check that none of the \emph{subdimensional} symmetries \eqref{def Q p} of the cubic lattice commute with $(\delta_2 \varphi)_f$ on triangular faces. This means that the dipole number, even on the cubic sublattice, is not generically conserved. Indeed, $(\delta_2 \varphi)_f$ on a triangular face serves to ``contract'' a dipole into a charged particle on the center of a cube. This way the dipole number and the total charge are both changed. At best, in some theories it will be possible to define a \emph{combination} of dipole number and total charge that is conserved.

It is thus necessary to be more specific in order to define a matter theory with subdimensional symmetries on the BCC lattice: the Hamiltonian cannot have the most general possible kinetic terms. This is similar, but not equivalent, to the need to exclude the four-sided tents (r.h.s.~of Fig.~\ref{fig BCC}) from the Hamiltonian of the pure $\Ff_2$ gauge theory in order to get a fracton theory.

The simplest way to preserve subdimensional symmetries on the BCC lattice is to construct the Hamiltonian using only pairs of kinetic operators $S_{f_1} S_{f_2}^{-1}$, with $f_1$ and $f_2$ being triangular plaquettes that form the ``wings'' of a tent shown on the l.h.s.~of Fig.~\ref{fig BCC}. This is in a sense a trivial solution: the product of such operators is equal to the kinetic operator $S_{f_3}$ on the square face that forms the third side of the tent in question. Thus this solution simply gives back the matter theory \eqref{def Lm} on the cubic lattice, with fields $\varphi\_c(\b r)$ decoupled from each other (at different points $\b r$) and from the fields $\varphi(\b r)$. Gauging the U(1) symmetry will keep these two matter fields decoupled, and the resulting theory would simply have matter on the cubic lattice interacting with $\Ff_2$ gauge fields on the BCC lattice. This is, in general, still different from a theory with both matter and $\Ff_2$ fields on the cubic lattice, and as such this is a novel theory that satisfies conditions F1--3.

It is also possible to construct theories in which matter and gauge fields both honestly live on the BCC lattice. The key is to include kinetic operators $S_f$ on only some triangular faces $f$ in the Hamiltonian, while allowing all possible square faces. For instance, consider only the triangular faces whose long edge is parallel to the $x$ direction. The corresponding operators $(\delta_2 \varphi)_f$ can contract any dipole on the cubic sublattice into a point charge --- as long as the dipole is parallel to the $x$ axis. Other dipoles are unaffected by these kinetic operators. In particular, this choice of dynamics makes sure that all excitations have restricted mobility, being able to move only in planes (just like in the original X-cube model).

Concretely, one Lagrangian that describes such a pure matter theory is\footnote{The couplings $e^2$ and $u^2$ do not generically need to appear at two different places in the Lagrangian. They do so only because the lattice Hamiltonian treats all plaquettes equally, but this is merely a choice that could be relaxed.}
\algns{
  \L\_m &= - \frac1{2u^2} \left[(\del_0\varphi)^2 + (\del_0\varphi\_c)^2\right] + \frac{e^2}2 \Big[\del_{ij}\varphi (\b r) \del^{ij}\varphi(\b r) + (\del_x \varphi(\b r) - \varphi\_c(\b r))^2 + \\
  & \qquad + (\del_x \varphi(\b r + \b e_y) - \varphi\_c(\b r))^2 + (\del_x \varphi(\b r + \b e_z) - \varphi\_c(\b r))^2 + (\del_x \varphi(\b r + \b e_y + \b e_z) - \varphi\_c(\b r))^2 \Big].
}
The $\del_{ij}\varphi \del^{ij}\varphi$ term is familiar from the cubic lattice, eq.~\eqref{def Lm}, and it comes from square plaquettes; a summation over $ij \in \{xy, yz, zx\}$ is implied, as usual. The other four terms come from all the terms of the form $S_f + S_{-f}$ on triangular plaquettes whose long edge is parallel to the $x$ axis. This theory has a subdimensional symmetry of the form \algns{\label{BCC symms}
  \varphi(\b r) &\mapsto \varphi(\b r) + \lambda(\b r), \\
  \varphi\_c(\b r) &\mapsto \varphi\_c(\b r) + \del_x \lambda(\b r),
}
for all functions $\lambda(\b r)$ that satisfy $\del_{ij} \lambda = 0$. Thus the space of symmetries is the same as in the simpler matter theory \eqref{def Lm} on the cubic lattice, and in particular condition F3 is met.

The fields $\varphi\_c$ can be integrated out to yield a (nonlocal) theory of fields $\varphi$ on the cubic lattice. This is one way to understand the particle content of this theory. In particular, it is easy to check that field configurations satisfying $\del_{ij} \varphi = 0$ all feature zero energy excitations: if $\del_x \varphi = \del_y \varphi = 0$, or if $\del_x \varphi = \del_z \varphi = 0$, the center fields $\varphi\_c$ are gapped out while $\varphi$ propagate along each $yz$ plane without a gap; and if $\del_x \varphi \neq 0$ while $\del_y \varphi = \del_z \varphi = 0$, then the linear combination $\varphi\_c - \del_x \varphi$ becomes gapped while $\varphi\_c + \del_x \varphi \equiv \theta$ becomes gapless with kinetic term $(1 + \frac{c}{p_x^2})(\del_0\theta)^2$. Thus the theory satisfies all the fracton requirements, F1--3. It is important to stress that this is achieved without conserving the dipole moment (in the $x$-direction) and the total charge: different symmetries \eqref{BCC symms} appear instead.

There are many other theories that can be built along these lines. A simple modification that preserves the Gaussianity of the theory (and all its subdimensional symmetries) is the addition of terms like $(\del_y \varphi\_c)^2 + (\del_z \varphi\_c)^2$. Further, it is possible to break dipole moment conservation in other directions too: all that is needed is to let some cubes have kinetic terms associated to triangular plaquettes with long sides parallel to $y$ or $z$. If all three possible choices appear on cubes throughout the lattice, no component of the dipole moment will be conserved, but subdimensional symmetries will still exist.

\subsection{Nonabelian $\Ff_2$ gauge theories}

A remarkable fact about $\Ff_2$ rank-two gauge theories in $(3+1)$D is that they can be defined for nonabelian gauge groups as well. Recall that ordinary two-form gauge theories cannot be given gauge-invariant field strengths in the nonabelian case: on a cubic lattice, the gauge fields live on faces, the field strengths on cubes, and it is impossible to find a combination of Lie group-valued variables on faces of the cube that is invariant under gauge transformations on all the links of the cube. However, in the $\Ff_2$ case, the field strengths live on belts, and their Lie group indices can be contracted in the same way as in a one-form nonabelian theory where field strengths are built out of operators on a one-cycle around a plaquette.

Concretely, pick a Lie group G. For every face $f$, consider the Hilbert space spanned by states $\qvec U_f$ for all $U \in \trm G$. The nonabelian generalizations of magnetic and electric operators $Z_f$ and $X_f$ from Subsection \ref{subsec pure F2} are operators $(Z_f)_{\alpha\beta}$ and $X^V_f$ that live on oriented plaquettes $f$ and act as
\gathl{\label{def X Z nonabel}
  (Z_f)_{\alpha\beta} \qvec U_f = U_{\alpha\beta} \qvec U_f, \quad   X_f^V \qvec U_f = \qvec{V U}_f.
}
Here, $\alpha$ and $\beta$ are ``color'' indices in the fundamental representation, while $V$ is an arbitrary element of G.\footnote{It is possible to define magnetic operators $(Z_f^R)_{\greek{ab}}$ in any representation $R$ of G, such that their action is $(Z_f^R)_{\greek{ab}} \qvec U_f = R_{\greek{ab}}(U) \qvec U_f$, with $\greek a, \greek b = 1, \ldots, \trm{dim} R$. However, any representation can be expressed as an appropriate linear combination of multiple fundamental (and antifundamental) ones, so working with just the operators $(Z_f)_{\alpha\beta}$ defined in eq.~\eqref{def X Z nonabel} is sufficient to cover all other representations.} On plaquettes with the opposite orientation, these operators act as
\bel{
  (Z_{-f})_{\alpha\beta} \qvec U_f \equiv \big(Z_f^{-1}\big)_{\alpha\beta} \qvec U_f = \left(U^{-1}\right)_{\alpha\beta} \qvec U_f, \quad X_{-f}^V \qvec U_f = \qvec{U V^{-1}}_f.
}

The Gauss operators that define a nonabelian $\Ff_2$ theory are
\bel{
  G_v^V \equiv \prod_{f \in \del_{-2} v} X^V_f.
}
The corresponding gauge-invariant field strengths are
\bel{\label{def tr W}
  \Tr\, W_b \equiv \Tr \prod_{f \in b} Z_f,
}
with matrix multiplication of the $(Z_f)_{\alpha\beta}$ understood in the product, and with the trace going over color indices. These operators are defined on every belt $b \subset c$ that satisfies $\del_2 b = 0$.

It may be instructive to demonstrate the action of Gauss operators $G_v^V$ in detail. The above definition works on any (orientable) lattice, but consider the cubic lattice for simplicity. A gauge transformation by $V \in \trm G$ at a site $v$ acts on a magnetic operator by conjugation,
\bel{
  (Z_f)_{\alpha\beta} \mapsto \left(G_v^V \right)^{-1} (Z_f)_{\alpha\beta} G_v^V.
}
If $v$ appears in $\del_2 f$ with coefficient $+1$, the gauge transformation is
\bel{\label{U transf left}
  (Z_f)_{\alpha\beta} \mapsto \left(X_f^V \right)^{-1} (Z_f)_{\alpha\beta} X_f^V = (V Z_f)_{\alpha\beta} = V_{\alpha\gamma} (Z_f)_{\gamma\beta},
}
where a sum over repeated color indices (in this case $\gamma$) is understood. This equality can be verified by computing the action of the gauge-transformed operator on a state $\qvec U_f$,
\bel{
  \left(X_f^V \right)^{-1} (Z_f)_{\alpha\beta} X_f^V \qvec U_f = \left( X_f^V \right)^{-1} (Z_f)_{\alpha\beta} \qvec{V U}_f = (VU)_{\alpha\beta} \qvec U_f.
}
Conversely, if $v$ appears in $\del_2 f$ with coefficient $-1$, the gauge transformation of $(Z_f)_{\alpha\beta}$ is
\bel{\label{U transf right}
  (Z_f)_{\alpha\beta} \mapsto \left(X_{-f}^V \right)^{-1} (Z_f)_{\alpha\beta} X_{-f}^V = \left(Z_fV^{-1} \right)_{\alpha\beta}.
}
If $v \in \del_2 f$ with any other nonzero coefficient $n$, $V$ should just be raised to the power of $|n|$ in eq.~\eqref{U transf left} (if $n > 0$), or in eq.~\eqref{U transf right} (if $n < 0$). Of course, if $v \notin \del_2 f$, $(Z_f)_{\alpha\beta}$ remains unchanged.

The electric operators $X_f^V$ also change under gauge transformations. To get gauge-invariant quantities out of them, it is necessary to build Casimir invariants of G by considering $X_f^V$'s for group elements $V$ that are infinitesimally close to the identity. This is a standard procedure and will not be reviewed here \cite{Kogut:1974ag}. 

It is also important to note that Gauss operators $G_v^V$ on different sites do not necessarily commute because they may act on a plaquette state $\qvec U_f$ by multiplying $U$ from the same side; this situation is not encountered in the usual $\Ff_1$ nonabelian theory, where degrees of freedom live on links and the gauge transformations on endpoints of a link necessarily multiply the group element from different sides. This new feature of $\Ff_2$ teories does not lead to an inconsistency, because it is still possible to define a gauge-invariant Hilbert space, i.e.~a set of states that are simultaneous eigenstates of all $G_v^V$'s with unit eigenvalue. These states are obtained by acting on the gauge-invariant state with zero electric flux,
\bel{
  \bigotimes_{f \subset \Mbb} \int\_G \d U \qvec U_f,
}
with the operators $\Tr \, W_b$ from \eqref{def tr W}.
%Gauss operators on the same site but with different $V$'s need not commute

It is straightforward to verify that $\Tr\, W_b$ is indeed gauge-invariant. Consider a belt $b$ winding along the $z$-axis on a cube with orientations shown on Fig.~\ref{fig field strengths}. The signs in the corners of each face $f$ indicate the sign with which the appropriate vertex figures in $\del_2 f$. As a two-chain, this belt is defined as the sum of four faces of this cube that are parallel to the $z$-axis,
\bel{
  b = f\_{front} + f\_{right} - f\_{back} - f\_{left}.
}
The signs are induced by the choice of orientation in Fig.~\ref{fig field strengths}, and they make sure the field strength
\bel{
  \Tr \, W_b = \Tr \prod_{f \in b} Z_f = \left(Z_{f\_{front}}\right)_{\alpha\beta} \left(Z_{f\_{right}}\right)_{\beta\gamma} \big(Z_{f\_{back}}^{-1}\big)_{\gamma\delta} \big(Z_{f\_{left}}^{-1}\big)_{\delta\alpha}
}
is gauge-invariant. A gauge transformation $G_v^V$ at any vertex $v \subset b$ causes two adjacent $Z$'s in the above product to transform according to eqs.~\eqref{U transf left} and \eqref{U transf right}, with the gauge parameters $V$ and $V^{-1}$ cancelling each other out. An analogous calculation can be carried out to verify the gauge invariance of operators $\Tr\, W_b$ for all other belts $b$.

%This also holds for belts that include more than four plaquettes. An important feature of nonabelian theories is that an operator $\Tr\, W_b$ over a large belt $b$ is not necessarily decomposable into products of operators $\Tr\, W_{b'}$ over smaller belts $b'$. In particular, if the rank of the gauge group is sent to infinity (the so-called planar or ``large $N$'' limit), all single-trace operators $\Tr\, W_b$  become independent of each other.

The formal steps above serve to establish gauge invariance directly on the lattice. The only step that remains in outlining these theories is to give their continuum actions. The analysis of their dynamics will not be performed here.

Assuming small fluctuations of gauge fields on a cubic lattice and passing to the continuum notation using $Z_f \equiv \e^{\i A_{ij}^a T^a}$, the gauge transformations are realized in the perfectly familiar way,
\bel{
  A_{ij}^a \mapsto A_{ij}^a + \del_{ij} \lambda^a + f^{abc} A_{ij}^b \lambda^c.
}
Similarly, the field strengths are the natural generalizations of \eqref{def B},
\bel{
  B_x^a  \equiv \del_y A_{zx}^a - \del_z A_{xy}^a + f^{abc} A_{zx}^b A_{xy}^c,
}
and so on. Their exponentials give the Wilson loops shown above, $W^i \equiv \e^{\i B_i^a T^a}$, and their traces are gauge invariant objects that naturally enter the Hamiltonian via terms like $\Tr(2 - W_b^i - W_{-b}^i)$. The simplest possible action of the pure gauge theory can be written in complete analogy with \eqref{def Lg},
\bel{
  \L = \frac1{2g^2} \Tr F_{\mu\nu} F^{\mu\nu}, \quad \mu\nu \in \{xy, yz, zx, 0x, 0y, 0z\}.
}

In short, the theory is defined in the most natural possible way, but tell-tale signs abound to show that this is not a usual theory. As in the Abelian case, $A_{ij}^a$ cannot be understood as a differential form.  The structure coefficients $f^{abc}$ are antisymmetric, and assuming symmetry of indices in $A_{ij}^a$ would kill a covariantly written $fAA$ term in the field strength. Many other aspects, however, remain unchanged compared to the ordinary one-form case. For instance, this is not a free theory, although at weak enough coupling, a perturbative fracton regime can be found. It would be extremely interesting to study its infrared dynamics and symmetries. This is left to future work.

\section{Concluding remarks} \label{sec outlook}

This paper examined in some detail the kinematics and rudimentary dynamics of four different models in $(3 + 1)$D that satisfy requirements F1--3. Here is a quick summary of this analysis:
\begin{enumerate}
  \item The X-cube models \eqref{Hg} have $\Z_2$ degrees of freedom on links of a cubic lattice, with three gauge constraints \eqref{Xcube G} per site. They have two distinct phases: confining ($g \rar \infty$) and topological ($g \rar 0$). In the confining phase, there is a unique ground state, and there are gapped glueball excitations with unrestricted mobility. In the topological phase, there is an exponential ground state degeneracy on a three-torus due to electric fluxes along the noncontractible cycles, and the excitations (fractons) are gapped, with restricted mobility. In both phases there is a kinematic, partly topological, two-form electric symmetry \eqref{2form symms}.
  \item The $\Z_2$ rank-two gauge theory of type $\Ff_2$ has one gauge constraint \eqref{def F2 G} on each site. With the Kogut-Susskind Hamiltonian (eq.~\eqref{def Hg} with $K = 2$), the theory also has a confining and topological phase. The confining phase has a unique ground state and three species of gapped glueballs without mobility restrictions, created by operators $W_c^i$ (eq.~\eqref{def W mu}). The topological phase has an exponentially degenerate ground state and gapped excitations with restricted mobility. In this case there is a kinematic, topological (in $\del_2$ homology), two-form electric symmetry \eqref{def Ue}.
  \item The U(1) rank-two gauge theory of type $\Ff_2$ in the Coulomb regime \eqref{def H Coulomb} has a continuum description \eqref{def H Coulomb cont}. Perturbatively, it has an exponential ground state degeneracy with gapless excitations \eqref{photon disp}, some of which have limited mobility. The theory has an electric two-form  symmetry that generalizes the one in the $\Z_2$ case (eq.~\eqref{def Ue}). In the continuum, there is also an exponential number of independent magnetic two-form symmetries.
  \item The free compact scalar theory \eqref{def Lm} naturally couples to the $\Ff_2$ gauge theory from the previous bullet point. It has an exponential ground state degeneracy, gapless excitations with limited mobility, and an exponential number of ``global'' symmetries \eqref{varphi transf}.
\end{enumerate}
In addition to these examples, explicit examples were given of theories on the BCC lattice and with a general Lie group as a gauge group. These illustrate the potential for novel behavior contained in the $\Ff_p$ class of theories. A particularly interesting lesson about fracton physics is that properties F1--3 can be satisfied even when charge and dipole moment are not conserved.

Taking a broad view, one question arises in its immediacy: what other $\Ff_p$ theories can there be in (3+1)D or below? This question can be answered by systematically checking all cases.

In two spatial dimensions, all gauge theories of type $\Ff_p$ have $p \leq 2$. The gauge fields of the $\Ff_2$ theory must live on faces, i.e.~they must have rank two. The gauge theory is necessarily topological, because no local gauge-invariant field strengths can be defined. The only gauge-invariant operators are electric fields $X_f$ and belts $W_b$ that wind around the spatial two-torus. The situation is superficially similar to an ordinary gauge theory in (1+1)D, except here not all electric fields $X_f$ are gauge-equivalent to each other. Thus, the theory is not quite trivial. However, it also does not have particularly interesting behavior: the theory is classical. This is easily seen from the fact that all local operators commute with each other, so every local Hamiltonian will be exactly diagonal in the $X_f$ eigenbasis.

In three spatial dimensions, it is possible to study two more $\Ff_p$ theories that have not been mentioned yet. First, there are $\Ff_3$ theories in which the gauge fields have rank three, and Gauss operators are defined on vertices. Second, there are $\Ff_2$ theories with rank-three gauge fields and Gauss operators defined on links. These suffer the same issues as the $\Ff_2$ theories in (2+1)D: they have no local gauge-invariant field strengths, and hence all local Hamiltonians are necessarily sums of commuting operators.

More interesting behavior will be found in higher dimensions. One step forward in this direction was taken in Ref.\ \cite{Li:2019tje}, whose generalization of X-cube to 4d dualizes to a $\Ff_2$ theory coupled to matter. This is the most obvious route for future research on this subject. The second most obvious route --- and a much more challenging one --- is to understand the dynamics of the theories discussed here. For instance, what is the critical coupling at which the $\Ff_2$ theory \eqref{Hg} deconfines?  Are the Higgs and confinement phases of the $\Ff_2$ theory coupled to fundamental matter smoothly connected, like in the $\Ff_1$ case \cite{Fradkin:1978dv, Banks:1979fi}?  Could there be continuous phase transitions in these theories, and can there be enhanced symmetry --- an analog of conformal symmetry --- at such critical points? What is the dynamics of nonabelian $\Ff_2$ theories at large $N$? Could it be possible that partition functions of these theories calculate new kinds of manifold invariants?

Then, there is also the question of the mathematical significance of the operators $\del_p$ upon which the $\Ff_p$ theories are based. They defy many of the intuitive features of differential geometry, but they may still prove perfectly consistent. Do they lead to generalized notions of Stiefel-Whitney classes, differential chain complexes, fiber bundles, and so on?

Finally, not all known fracton theories can be interpreted as gauge theories of $\Ff_p$ type. The theories discussed here are all ``type I'' fracton theories \cite{Vijay:2016phm}. Haah's code \cite{Haah_2011, Tian:2018} is an example of a ``type II'' theory not mentioned here. Perhaps there exists an even more general class of gauge theories that would contain all fracton theories.

\section*{Acknowledgments}

It is a pleasure to thank Maissam Barkeshli, Mike Hermele, Alex Kubica, Steve Shenker, and Kevin Slagle for insightful discussions. This work was initiated at the Aspen Center for Physics, which is supported by National Science Foundation grant PHY-1607611. Part of this work was done at the Perimeter Institute for Theoretical Physics, which is supported by the Government of Canada through Industry Canada and by the Province of Ontario through the Ministry of Economic Development \& Innovation. The remaining part of this work was done with support from the Simons Foundation through \emph{It from Qubit: Simons Collaboration on Quantum Fields, Gravity, and Information}, and from the Department of Energy Office of High-Energy Physics through Award DE-SC0009987.

\bibliographystyle{ssg}
\bibliography{XCubeRefs}

\end{document}